\documentclass[11pt]{article}

\usepackage[margin=1in]{geometry}
\usepackage[T1]{fontenc}
\usepackage[utf8]{inputenc}
\usepackage{amsmath,amssymb,amsthm}
\usepackage{graphicx}
\usepackage{booktabs}
\usepackage{array}
\usepackage{float}
\usepackage{natbib}
\usepackage{setspace}
\usepackage[colorlinks=true,citecolor=blue,linkcolor=blue,urlcolor=blue]{hyperref}
\usepackage{pdfpages}

\graphicspath{{./}}
\newtheorem{principle}{Principle}

\title{Spending Scarce Confirmatory PET Measurements: Target-Aligned Validation in A4/LEARN}
\author{Elvis Han Cui\textsuperscript{1,2}\\
Qiang Yang\textsuperscript{3}\\
Meredith Mengmeng Zhang\textsuperscript{3}\\[0.5em]
\small \textsuperscript{1}Clinical Functional Service Provider (cFSP) Department, Kuntuo, an IQVIA company\\
\small \textsuperscript{2}Department of Biostatistics, University of California, Los Angeles\\
\small \textsuperscript{3}Department of Biometrics, Hengrui Pharmaceuticals\\[0.25em]
\small Elvis Han Cui's Kuntuo/IQVIA affiliation is listed as the affiliation\\
\small at the time this work was conducted.\\[0.25em]
\small Corresponding author: Elvis Han Cui, \href{mailto:elviscuihan@g.ucla.edu}{elviscuihan@g.ucla.edu}\\
\small Author email: Qiang Yang, \href{mailto:yq88627@163.com}{yq88627@163.com}\\
\small Author email: Meredith Mengmeng Zhang, \href{mailto:meredith.zhang.cdm@gmail.com}{meredith.zhang.cdm@gmail.com}}
\date{September 9, 2026}

\begin{document}
\maketitle

\begin{abstract}
Anti-amyloid therapies and blood-based biomarkers are changing Alzheimer disease workups into a two-stage measurement workflow: screen broadly with cheaper information, then spend scarce confirmatory amyloid measurements where they support the decision that will be reported. Amyloid positron-emission tomography (PET) remains one such protocol measurement for amyloid burden, but PET slots, trial budgets, and payer-facing evidence packages are finite. This paper asks a deliberately operational question: when is simple transparent PET validation enough, and when is a fitted residual-uncertainty score worth the added complexity? For a weighted protocol target, the first-order value of validating subject \(i\) is the product of target influence and residual protocol uncertainty. Generic uncertainty sampling uses only the second factor and can spend PET measurements on subjects that are hard to predict but weak for the scientific, clinical, or commercial claim. We apply this rule to the A4/LEARN PET archive, treating observed PET as a design laboratory for scarce-confirmation studies. For the primary APOE4 carrier versus non-carrier contrast in Centiloid 24-or-higher PET positivity, simple APOE4-balanced validation recovers nearly all of the target-specific gain: at PET budget 200, the confidence-interval width ratio relative to random validation is 0.923 for APOE4 balancing and 0.914 for target-specific scoring, while generic uncertainty sampling is 0.980. Other targets behave differently: target-specific scoring gives larger gains for an age-slope analysis and for cutoff-indexed PET positivity. The practical message is simple: spend scarce protocol measurements according to the claim being validated, not only according to prediction uncertainty.
\end{abstract}

\noindent\textbf{Keywords:} amyloid PET; active sampling; design-based inference; two-phase validation; APOE4.

\section{Why scarce confirmatory measurement is the problem}

Validation studies often begin with an awkward asymmetry. A large archive contains variables that are cheap enough to collect broadly, while the measurement that defines the scientific endpoint is expensive, invasive, slow, or restricted by assay capacity. The operational question is not merely whether a surrogate model predicts well. It is which expensive confirmations should be bought, scanned, assayed, or adjudicated when the validation budget is smaller than the screened population.

This question has become commercially and clinically sharper in Alzheimer disease. Anti-amyloid therapies require evidence of amyloid pathology before treatment, making biomarker confirmation part of the treatment workflow rather than only a research endpoint \citep{FDA2026LeqembiLabel,FDA2024KisunlaLabel}. At the same time, blood-based biomarkers and other lower-cost measures are expanding the front end of the diagnostic funnel \citep{PalmqvistEtAl2024BloodBiomarkers,PalmqvistEtAl2025BBMGuideline}. Expanded coverage and broader screening do not eliminate the PET bottleneck; they make the allocation problem more visible \citep{CMS2023AmyloidPET}. A diagnostic company, trial sponsor, imaging network, or payer-facing evidence team may care about an APOE4 contrast, an age trend, a cutoff choice, or a treatment-eligibility rule. Those are different targets, and they need not be served by the same validation subset.

The same target-measurement tension appears in public Hengrui development examples, where the claim may be an all-comer treatment effect, a biomarker-positive subgroup, or an assay-intensive exploratory signal. In the CameL phase 3 lung-cancer trial, the protocol distinguished progression-free survival in all patients from progression-free survival in PD-L1-positive patients; in CAPTAIN, PD-L1 and MHC-I/MHC-II measurements were analyzed as response-selection biomarkers; and in a camrelizumab-plus-apatinib triple-negative breast cancer analysis, tumor and blood immune measurements were explicitly treated as exploratory biomarker readouts \citep{ZhouEtAl2021CameL,YangEtAl2021CAPTAIN,LiuEtAl2021TNBCBiomarkers}. A Hengrui Department of Biometrics population pharmacokinetic/pharmacodynamic analysis illustrates the same operational logic in a different measurement setting: model-based evidence linked protocol endpoints such as glucose-infusion rate and HbA1c to dose-ratio decisions \citep{TangEtAl2026INS068PopPKPD}. These examples are not empirical inputs to the A4/LEARN analysis; they motivate the general design point that scarce, assay-intensive, or adjudicated measurements should be prioritized according to the claim they are meant to support.

Amyloid PET in A4/LEARN makes the distinction concrete. The Anti-Amyloid Treatment in Asymptomatic Alzheimer's Disease study and the companion LEARN study contain genetic, demographic, cognitive, plasma, and magnetic-resonance imaging information, together with amyloid PET on a large subset of participants \citep{SperlingEtAl2020A4,SperlingEtAl2023Solanezumab}. PET is costly but scientifically central: it is the protocol measurement for amyloid burden, summarized here as Centiloid 24-or-higher positivity \citep{ClarkEtAl2012Florbetapir,KlunkEtAl2015Centiloid,NavitskyEtAl2018FlorbetapirCentiloid,JackEtAl2024RevisedCriteria}. In a future study, investigators might screen broadly with cheaper information and confirm PET on only a subset. Which subset should be validated?

The answer depends on the question the PET measurements are meant to answer. The tempting response is to build the best possible PET-risk model and validate subjects whose outcomes are most uncertain. That response is natural if the business objective is to improve the predictor. It is not always natural if the objective is to support a prespecified claim, contrast, or regulatory-style evidence package. This paper focuses on the APOE epsilon-4 (APOE4) carrier versus non-carrier contrast in PET positivity. APOE4 is measured before PET, is biologically established as an Alzheimer disease risk factor, and defines a clinically interpretable target contrast \citep{FarrerEtAl1997APOEJAMA,SperlingEtAl2020A4}. For this target, a transparent APOE4-balanced PET validation design performs almost as well as a fitted target-specific design. That is the main empirical lesson. The secondary lesson is just as important: the same conclusion does not automatically transfer to an age slope, a cutoff-indexed PET endpoint, or a cohort-structured diagnostic contrast.

The contribution is therefore not another claim that a surrogate model predicts PET well. The contribution is a decision rule for using surrogate information without letting the surrogate redefine the target. The rule is old in spirit, drawing on two-phase sampling, model-assisted survey estimation, and missing-by-design inference \citep{Neyman1934,Cochran1977SamplingTechniques,SarndalEtAl1992ModelAssisted,RobinsRotnitzkyZhao1994,BreslowEtAl2009,GilbertYuRotnitzky2014}. The reframing for this application is operational: define the claim, identify the protocol measurement that anchors it, record validation probabilities before outcome use, and spend scarce measurements where they reduce uncertainty for that claim. The paper can be read in four steps: Section 2 defines the A4/LEARN scarce-PET decision, Section 3 gives the recorded-probability rule, Section 4 shows when simple APOE4 balancing is enough, and Sections 5--6 show when target-specific scoring or pilot learning becomes more useful.

\section{The A4/LEARN validation decision}

The analysis uses the July 2024 A4/LEARN controlled-access release. The processed first-phase archive contains 6,945 rows; 4,492 have nonmissing Centiloid PET; and 4,460 have both PET and APOE4 status. Individual-level data are not redistributed. The complete PET archive is used retrospectively as a design laboratory: hide PET outcomes, spend an emulated PET validation budget according to candidate rules, and evaluate finite-archive PET inference for prespecified targets.

The primary endpoint is
\[
Y_i=1\{C_i\ge 24\},
\]
where \(C_i\) is Centiloid PET. CL24 is used as an operational florbetapir/Centiloid protocol threshold, not as a universal disease boundary. The primary target is the finite-archive APOE4 carrier versus non-carrier difference in CL24 positivity over the observed design-emulation archive; population language is shorthand for this empirical contrast, not an external-validity claim.

The surrogate score uses first-phase demographic, APOE4, cognitive, plasma, and MRI information where supported. Scores are built and evaluated under cross-fitting when PET outcomes are used to assess prediction, so an individual's own PET value is not fed back into its predicted risk. Cross-fitting, however, is not the same thing as prospective availability. Demographics, APOE4, and design variables have broad first-phase support; cognitive summaries are available for 1,708 PET+APOE participants with at least one listed variable and 1,167 with all listed variables; plasma summaries are available for 1,926 and 1,349 under the same convention; pTau217 has 1,439 and 1,000; and MRI summaries have 1,792 complete cases. The main claims therefore concern validation design under documented support, not clinical deployment of a particular PET-risk model.

The design comparison includes four practically distinct choices:
\[
\begin{array}{ll}
\text{Random validation} & \text{baseline design-based sampling;}\\
\text{Generic uncertainty} & \text{validate subjects with high PET prediction uncertainty;}\\
\text{APOE4-balanced validation} & \text{spend equal expected PET counts in carrier strata;}\\
\text{Target-specific validation} & \text{allocate by target influence times residual uncertainty.}
\end{array}
\]
All ratios below are relative to random validation at the same PET budget. Budget 200 is a readable mid-range scarce-PET scenario in a broader 50/100/200/400/800 budget curve, not a uniquely optimized choice. Oracle lower-bound rows are used only as infeasible diagnostics and are not proposed as deployable designs. The intervals are finite-archive design intervals for validation randomness; they do not add superpopulation sampling variation, surrogate-model selection uncertainty, future feature-availability uncertainty, or biological measurement-error components beyond the chosen PET protocol endpoint.

\begin{figure}[H]
\centering
\includegraphics[width=\textwidth]{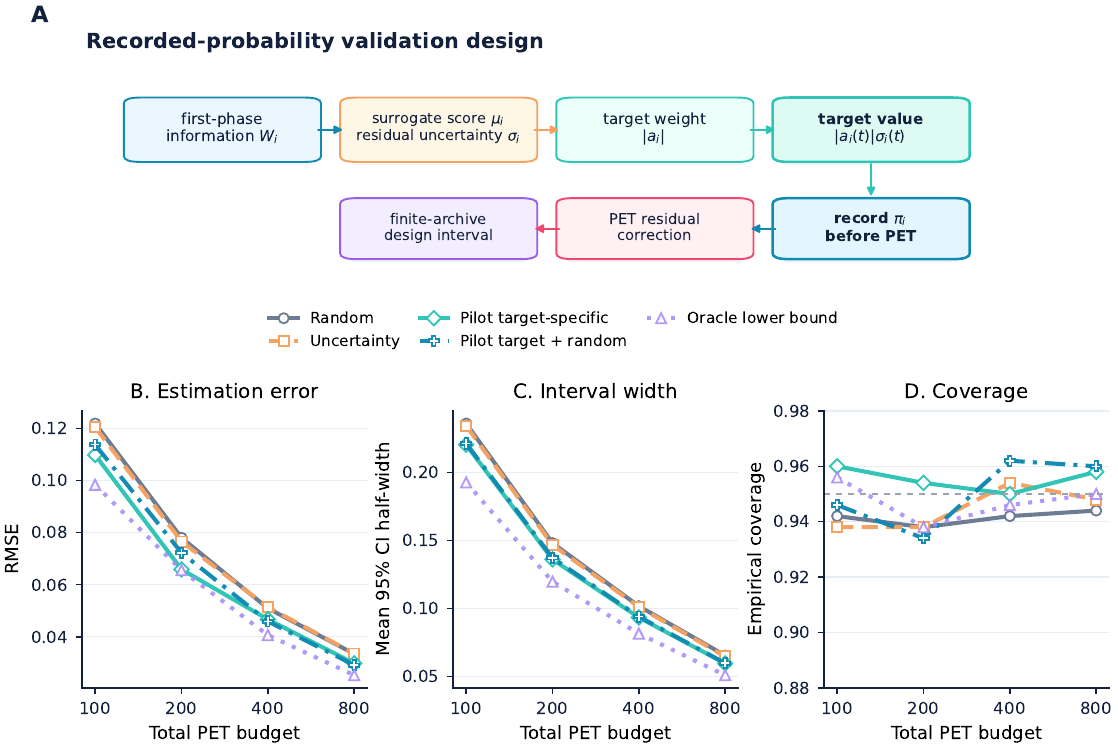}
\caption{Recorded-probability validation workflow. First-phase information supports prediction and residual-uncertainty scoring; validation probabilities are recorded before PET outcomes enter the final correction; inference reports PET-defined design intervals.}
\label{fig:workflow}
\end{figure}

\section{Target-aware rule and recorded estimator}

Let \(Y_i(t)\) be the protocol measurement for subject \(i\), indexed by threshold or target label \(t\), and let
\[
\theta(t)=\sum_i a_i(t)Y_i(t)
\]
be the finite-archive target. The weight \(a_i(t)\) encodes what the study is trying to estimate. In an APOE4 contrast, it separates carriers from non-carriers; in an age-slope analysis, it reflects age leverage; in a cutoff curve, it changes with the PET threshold.

Let \(\mu_i(t)\) be a first-phase prediction of the PET-defined quantity and let \(\sigma_i(t)\) describe residual protocol uncertainty after the first-phase information is used. A validation design records a positive probability \(\pi_i\) before PET outcomes enter the final correction. The augmented estimator is
\[
\widehat{\theta}(t)
=
\sum_i a_i(t)\mu_i(t)
+
\sum_i\frac{R_i}{\pi_i}a_i(t)\{Y_i(t)-\mu_i(t)\},
\]
where \(R_i\) indicates validation.

\begin{principle}[Target-aligned validation value]
For a scalar target, the first-order value of validating subject \(i\) is proportional to
\[
|a_i(t)|\sigma_i(t).
\]
For an indexed target, replace this by a prespecified summary such as
\[
S_i=\left\{\sum_t \omega_t a_i(t)^2\sigma_i(t)^2\right\}^{1/2}.
\]
\end{principle}

This is the design message. Generic uncertainty sampling uses \(\sigma_i(t)\). Target-aligned validation uses \(|a_i(t)|\sigma_i(t)\). The difference matters whenever prediction uncertainty and target influence point to different subjects. Conditional on the archive, the first-phase scores, and the recorded probabilities,
\[
E_R\left[
\sum_i \left(\frac{R_i}{\pi_i}-1\right)a_i(t)\{Y_i(t)-\mu_i(t)\}
\right]=0.
\]
Thus the surrogate model can improve precision, but correctness of the surrogate model is not what centers the estimator. Centering comes from the recorded validation probabilities and positivity. In a real deployment, \(S_i\) must be computed from first-phase, historical, cross-fitted, or pilot information available before the final PET validation decision. Full-PET residuals are used here only to evaluate candidate designs in the retrospective laboratory.

\section{Primary APOE4 result: simple balancing is nearly enough}

For the APOE4 contrast, the PET protocol contrast is 0.334 and the surrogate-only contrast is about 0.355. The cross-fitted surrogate AUC is about 0.779, but the AUC is not the design conclusion. The design conclusion is that uncertainty sampling adds little for this estimand, while APOE4 balancing recovers most of the target-specific gain.

At PET budget 200, the CI-width ratio is 0.980 for generic uncertainty sampling, 0.923 for APOE4-balanced validation, and 0.914 for target-specific scoring. In words, the fitted target-specific design is best among these rules, but the simple balanced design is almost indistinguishable for practical planning. Balanced validation is easier to explain, audit, and implement than a residual-uncertainty score, so it should be the serious default comparator for this target.

\begin{figure}[H]
\centering
\includegraphics[width=\textwidth]{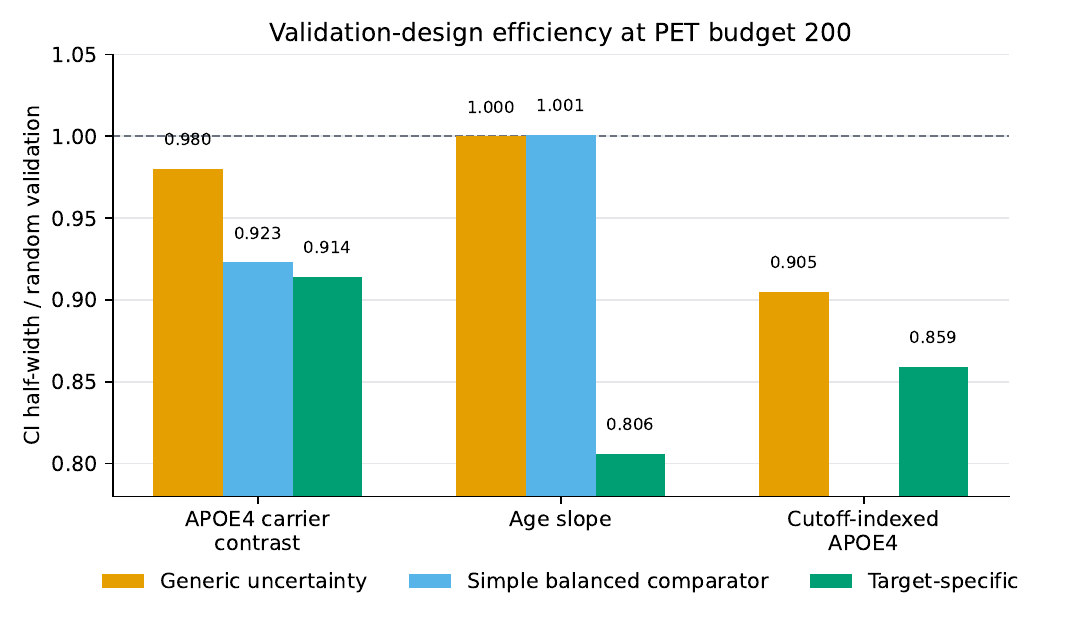}
\caption{Validation-design efficiency at PET budget 200. The APOE4 result is the primary planning result: simple APOE4 balancing nearly matches target-specific scoring. The age and cutoff rows show the scope condition: target-specific residual-uncertainty scoring is more useful when influence is heterogeneous or the target is indexed.}
\label{fig:design-ratios}
\end{figure}

This result should not be oversold. It does not say that fitted scores are unnecessary in every PET validation study. It says that, for this two-group target, the most important influence structure is already visible before modeling: APOE4 status defines the contrast. Once the design guarantees adequate PET measurement in both strata, the remaining target-specific residual-uncertainty refinement is modest.

A useful stress check removes APOE4 from the PET-risk surrogate while keeping APOE4 as the prespecified target-defining variable. That stricter score has AUC 0.724 and compresses the APOE4 surrogate-only contrast to 0.069 versus the PET contrast 0.334, giving bias -0.265. Target-specific validation still improves precision relative to random validation at budget 200, with CI-width ratio 0.952 and RMSE ratio 0.961; generic uncertainty sampling has CI-width ratio 1.004. This separates APOE4's role as an estimand-defining stratum from its optional role inside a prediction model.

\begin{table}[H]
\centering
\scriptsize
\caption{Design ratios at PET budget 200 for selected A4/LEARN targets. Ratios are relative to random validation; values below 1 indicate narrower intervals. One-dimensional balanced rows are checks, not a universal dominance ranking.}
\label{tab:design-ratios}
\begin{tabular}{p{0.35\textwidth}ccc}
\toprule
Target & Generic uncertainty & Balanced comparator & Target-specific \\
\midrule
APOE4 carrier contrast & 0.980--0.988 & 0.923 & 0.914--0.915 \\
Age slope & 1.000 & 1.001 & 0.806 \\
CL24 within cutoff-indexed APOE4 curve & 0.905 & -- & 0.859 \\
\bottomrule
\end{tabular}
\end{table}

\section{Scope checks: when targeting matters}

The APOE4 result is deliberately paired with scope checks. For the age slope in PET positivity, the surrogate-only point estimate is close to the PET protocol value, but target-specific validation still matters for precision: the CI-width ratio at budget 200 is 0.806, while generic uncertainty sampling is 1.000 and age-quartile balancing is 1.001. Coarse balancing is not the same as allocating by leverage and residual protocol uncertainty.

For cutoff-indexed APOE4 positivity over Centiloid thresholds 20 to 30, the target changes with the threshold. At the primary threshold \(c=24\), the PET contrast is 0.334 and the threshold-specific surrogate contrast is 0.347. A cutoff-range design gives a CI-width ratio of about 0.859, compared with about 0.905 for generic uncertainty validation. The endpoint is not a single-cutoff artifact: at Centiloid thresholds 20, 24, and 30, the PET APOE4 contrasts are 0.326, 0.334, and 0.318.

\begin{figure}[H]
\centering
\includegraphics[width=\textwidth]{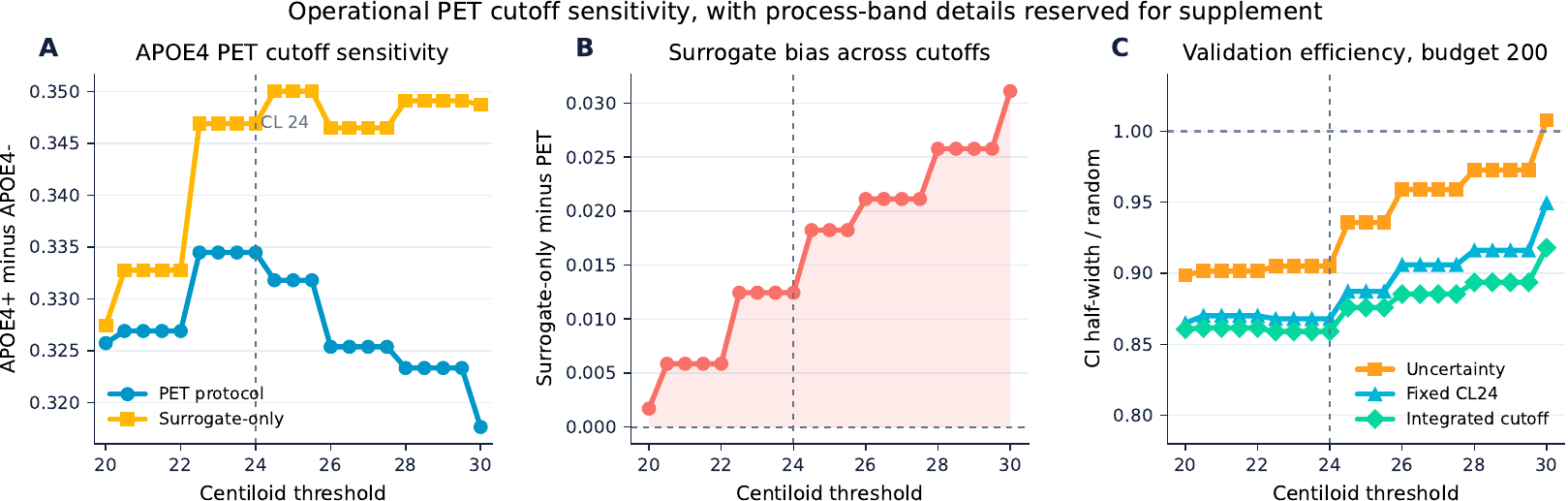}
\caption{APOE4 PET cutoff sensitivity across operational Centiloid positivity thresholds. The target-specific design is evaluated over the prespecified threshold range rather than only at CL24.}
\label{fig:cutoff}
\end{figure}

The A4-versus-LEARN comparison is moved to the diagnostic layer. It is not a causal or biological cohort contrast. Its value is to show that surrogate error can align with cohort structure: a cohort-unaware surrogate has AUC about 0.779 but gives a surrogate-only A4-versus-LEARN contrast of 0.380 compared with the PET protocol contrast of 0.873. Adding cohort information raises AUC to about 0.908 and reduces the contrast-aligned bias from -0.493 to -0.039. This stress test supports the target-aware warning but should not occupy the primary biomedical story.

\section{Deployable pilot evidence}

The retrospective archive can use all PET outcomes to evaluate designs, but a real study cannot. The deployable version is a two-wave design:
\[
\begin{aligned}
&\text{random PET pilot}
\longrightarrow
\text{learn residual uncertainty}\\
&\longrightarrow
\text{record target-aligned probabilities}\\
&\longrightarrow
\text{estimate with the recorded design.}
\end{aligned}
\]

In the existing A4/LEARN pilot emulation, a random pilot of 50 followed by a target-specific wave of 150 gives an RMSE ratio of 0.844 and a CI-width ratio of 0.919 relative to pilot-random validation, with coverage 0.954. At total budget 400 with a random pilot of 100, the corresponding ratios are 0.917 and 0.918, with coverage 0.950. These numbers are less dramatic than full-archive oracle diagnostics, but they are more relevant for practice because they do not require unobserved PET outcomes to choose the second wave.

The pilot design also clarifies the role of machine learning. A richer learner may reduce prediction error, but it is useful for validation only if it improves residual uncertainty in the influential part of the target and is available before the validation decision. In the model-ladder audit, flexible safe-feature scores support the same qualitative APOE4 conclusion, while the pTau217 sensitivity is kept outside the main design because its archive timing is not uniformly pre-PET.

\section{Discussion}

The paper's message is intentionally narrower than the earlier technical version. A good surrogate model is useful, but validation should be aligned with the target that PET is meant to estimate. In A4/LEARN, this distinction leads to a practical planning rule. For the primary APOE4 PET contrast, start with APOE4-balanced validation; it is transparent and nearly matches the target-specific design. For targets with uneven influence, cutoff-indexed structure, or within-stratum residual heterogeneity, move to target-aligned residual-uncertainty scoring. For prospective deployment, learn residual uncertainty from historical data or a random pilot and record validation probabilities before PET outcome use.

Several boundaries remain. The analysis is retrospective: the complete PET archive is used to evaluate scarce-validation designs, not to claim that future PET outcomes would be observed under the same mechanism. PET visual read and CL24 agree in 92.2\% of 4,492 paired rows, with Cohen's kappa 0.821, but visual read is a plausibility audit rather than a replacement endpoint. Plasma, MRI, pTau217, and cognitive features have heterogeneous timing and support, so they should be used for targeting only when available in the intended first-phase cohort. Crossed stratified grids and cutoff-aware balanced validation remain outside the current performance package.

Those limitations are also why the design convention matters. When expensive protocol measurements define the endpoint, the study should document the target, the first-phase support, the validation probabilities, and the final PET-defined estimator. Prediction can help decide where to measure, but the measurement design should remain answerable to the scientific estimand.

The same discipline is useful beyond this PET example, but it should not be forced into the main empirical claim. In biophysical chemistry and stochastic biochemical kinetics, the scientific object is often a kinetic, thermodynamic, or trajectory-level quantity, while cheaper proxy measurements may be available before a high-fidelity assay or imaging protocol is run. Work associated with Qian's stochastic thermodynamic view of biochemical systems is a useful conceptual reminder: the measured trajectory, state variable, flux, or protocol-defined functional has to be specified before auxiliary measurements are allowed to guide inference \citep{Qian2006OpenSystems,Qian2007PhosphorylationEnergy}. The present paper does not model biochemical reaction networks; it borrows only the measurement-design lesson that proxy information should guide protocol measurement, not replace the protocol quantity.

A separate mathematical extension would replace the binary PET target by a time-indexed or paired-time endpoint. Product-integral and bivariate survival functionals, including Dabrowska-type estimators on the plane, give a natural language for such targets \citep{Dabrowska1988KaplanMeierPlane,AndersenEtAl1993CountingProcesses,Hougaard2000MultivariateSurvival}. That extension is compatible with the recorded-probability contract, but it is not needed for the A4/LEARN empirical message. Keeping it as an extension makes the paper easier to read: the current claim is about scarce confirmatory PET measurements for scalar and cutoff-indexed amyloid targets.

\section*{Data availability}

Individual A4/LEARN participant-level data are controlled-access study data available to qualified researchers through the A4 and LEARN Study Data Portal, A4StudyData.org, subject to registration, approval, and portal data-use requirements. Controlled-access source files are not redistributed. The accompanying reproducibility materials provide manuscript sources, analysis and simulation code, cleared aggregate tables and figures, requirements information, and synthetic simulation checks.

\section*{Use of AI/NLP tools}

OpenAI Codex and ChatGPT were used as editorial and programming assistants for drafting, code organization, formatting checks, and submission preparation. The authors reviewed and take responsibility for all manuscript content, code, analyses, and conclusions.

\bibliographystyle{plainnat}
\bibliography{references_seed}

\clearpage


\section*{Supplementary material}

This full proof supplement provides the inferential support for the A4/LEARN recorded-probability validation analysis: comparator boundaries, complete design-based arguments, simulation calibration, empirical robustness checks, and compact reproducibility notes. Here A4/LEARN denotes the Anti-Amyloid Treatment in Asymptomatic Alzheimer's Disease study and its companion Longitudinal Evaluation of Amyloid Risk and Neurodegeneration study; PET denotes positron emission tomography; and APOE4 denotes apolipoprotein E epsilon-4 carrier status.

# Supplement A. Related Work and Comparator Boundaries

The manuscript combines classical validation-sampling ideas with modern prediction-assisted inference and active measurement acquisition. The augmented estimating equations are rooted in two-phase sampling and semiparametric missing-data theory, including missing-by-design regressors, validation samples, generalized Horvitz-Thompson correction, efficient observed-data influence functions, and optimal auxiliary-covariate-based phase-two designs (Bickel et al., 1993; Robins et al., 1994; van der Vaart, 1998; Tsiatis, 2006; Breslow et al., 2009; Gilbert et al., 2014). In this paper that literature supplies the inferential backbone for a finite-archive PET validation problem.

Prediction-powered and post-prediction inference show how many predictions and a smaller set of observed labels can support valid estimation after debiasing (Wang et al., 2020; Angelopoulos et al., 2023a,b; Zrnic and Candes, 2024b). A4/LEARN changes the emphasis from correcting a fixed labeled sample to deciding which scarce protocol measurements should be acquired for a registered scientific contrast.

Active statistical inference, optimized labeling, MultiPPI, AM-PPI, robust ASI, Active Sampling, and optimal subsampling are the closest allocation relatives because they place label or measurement acquisition inside an inferential task (Zrnic and Candes, 2024a; Li

et al., 2025; Ma and Candes, 2026; Cowen-Breen et al., 2026; Brawand et al., 2026; Imberg et al., 2025; Fithian and Hastie, 2014; Wang et al., 2018, 2019). We treat those papers as complementary: they provide the general acquisition-for-inference language, while the A4/LEARN case study fixes the protocol measurement, first-phase support, PET target, and recorded-probability interval convention.

Active learning and uncertainty sampling motivate a practical baseline of labeling prediction-ambiguous cases (Lewis and Gale, 1994; Settles, 2009). That objective is different from estimating a contrast. In the A4/LEARN design, a unit is valuable when it is both influential for the target and uncertain after the first-phase surrogate; generic prediction ambiguity alone can point elsewhere.

Measurement-error and misclassification literatures explain why high predictive performance can still leave estimand-aligned bias (Copeland et al., 1977; Carroll et al., 2006; Dawid and Skene, 1979). Amyloid PET is not treated as metaphysical truth; it is the prespecified protocol measurement that defines the analysis target. The Alzheimer's biomarker literature motivates the particular protocol setting: A4 screening, Centiloid standardization, florbetapir PET validation and calibration, and blood-biomarker measurement all make scarce PET confirmation a realistic design problem (Sperling et al., 2020; Clark et al., 2012; Klunk et al., 2015; Navitsky et al., 2018; Palmqvist et al., 2024; Jack et al., 2024).

# Supplement B. Theory and Complete Proofs

The results in this section formalize the finite-archive validation experiment used in the A4/LEARN analysis. Sections B.1-B.4 give recorded-probability validity, variance, and adaptive-design support. Section B.4a records the corresponding superpopulation efficient-information interpretation. Sections B.5-B.6 cover the cutoff-indexed process and first-order allocation calculations used by the Centiloid sensitivity analysis. Sections B.7-B.9 record efficiency diagnostics and the variance convention for reported intervals. Section B.10 gives

the product-integral bivariate survival extension, and Section B.11 records boundaries for nonlinear targets.

*Standing design convention.* All probability and expectation statements in Supplement B are design-based. They condition on the finite analysis archive through the sigma-field $\mathcal{A}_n$, while probability assignment itself uses only the deployable history $\mathcal{D}_k$ available before the $k$th validation draw. The proof filtration is

$$\mathcal{G}_k = \mathcal{A}_n \vee \sigma(\pi_j, R_j : j \leq k),$$

with $\pi_k$ recorded before $R_k$. For martingale increments, conditioning is on the pre-draw field

$$\mathcal{G}_{i-1}^{+} = \mathcal{G}_{i-1} \vee \sigma(\pi_i),$$

so the assigned probability is known before $R_i$ is drawn while $\mathcal{A}_n$ fixes the finite-archive residual contribution. The target is finite-archive and protocol-defined.

## Design Convention and Proof Setup

The mathematical point of the supplement is modest but important: once the target, first-phase information, validation probabilities, and protocol measurement are fixed, the validation indicators generate the remaining design randomness. The recorded-probability residual correction is then a centered triangular array for fixed designs and a martingale array for pilot or adaptive designs. The cutoff analysis applies the same argument to a cutoff-indexed process. The allocation calculations are classical first-order variance programs for these recorded validation values.

All random variables are defined on a probability space $(\Omega, \mathcal{F}, \mathbb{P})$. The sigma-field $\mathcal{A}_n$ is generated by the finite analysis archive, including the first-phase variables, target weights, protocol outcomes, and fixed nuisance or control-variate quantities used to define the finite-

archive target. The deployable history $\mathcal{D}_k$ contains the information available when assigning the $k$th validation probability: first-phase data, previous validation indicators, and previous validated protocol measurements, but not unobserved future protocol outcomes. A probability assignment is valid for the design proof only when $\pi_k$ is $\mathcal{D}_{k-1}$-measurable and recorded before $R_k$ is drawn. After conditioning on $\mathcal{A}_n$, residuals involving the full protocol outcomes are fixed; the martingale randomness is carried by the validation indicators under $\mathcal{G}_k$.

## B.1 Setup and Conditioning

Let $i = 1, \ldots, n$ index the first-phase cohort. The first-phase data are $W_i$, which include baseline covariates, scientific contrast variables, surrogate measurements, and design-relevant metadata. The protocol measurement $Y_i$ is observed by the analyst only when $R_i = 1$. For proof purposes, $\mathcal{A}_n$ conditions on the complete finite archive; for design purposes, the probability $\pi_i$ is assigned from $\mathcal{D}_{i-1}$, recorded, and only then is $R_i$ realized.

The full-protocol estimating equation is

$$U_n(\theta) = n^{-1} \sum_{i=1}^{n} \psi(Y_i, Z_i; \theta),$$

where $Z_i$ contains first-phase variables needed in the estimating function. The finite-population target $\theta_n$ solves $U_n(\theta_n) = 0$.

The augmented two-phase estimating equation is

$$\widehat{U}_n(\theta) = n^{-1} \sum_{i=1}^{n} \left[ m_i(\theta) + \frac{R_i}{\pi_i} \{\psi(Y_i, Z_i; \theta) - m_i(\theta)\} \right],$$

where $m_i(\theta) = m(W_i, Z_i; \theta)$ is a first-phase regression, prediction, or control variate. Linear contrasts correspond to $\psi_i(\theta) = a_i Y_i - \theta$ after centering or to the equivalent direct contrast estimator.

If $m_i$ is learned from validation outcomes, the learning step is separated by sample splitting, cross-fitting, historical information, or a pilot/main split. The final correction then

uses a control variate fixed at the time the corresponding probability is assigned. Surrogate and pilot models can change efficiency through $\pi_i$ and through residual size; they do not supply the centering identity, which comes from recorded predictable probabilities.

**Assumption set B.1 (recorded-probability design).**

(A1) *Finite archive and target.* Conditional on $\mathcal{A}_n$, the target is

$$\theta_n(t) = n^{-1} \sum_{i=1}^{n} \psi_t(Y_i, W_i), \qquad t \in T,$$

or, for linear contrasts, $\theta_n(t) = \sum_i a_i(t) Y_i(t)$.

(A2) *Recorded predictable design.* Validation indicators satisfy

$$\mathbb{P}(R_i = 1 \mid \mathcal{D}_{i-1}) = \pi_i,$$

where $\pi_i$ is $\mathcal{D}_{i-1}$-measurable and recorded before the corresponding protocol outcome contributes to estimation. In proofs the residual increments are adapted to $\mathcal{G}_i$, with the predictable pre-draw field $\mathcal{G}_{i-1}^{+} = \mathcal{G}_{i-1} \vee \sigma(\pi_i)$.

(A3) *Positivity and tail control.* $0 < \pi_i \leq 1$. For normal approximations, the largest inverse-probability residual contribution is controlled by the stated Lindeberg condition.

(A4) *Control variate.* The first-phase score $m_i(\theta, t)$ or $\mu_i(t)$ is fixed conditional on the design history, or is learned using cross-fitting, sample splitting, historical data, or pilot/main separation.

(A5) *Smooth Z-estimation.* $U_n(\theta, t)$ is differentiable near $\theta_n(t)$, with nonsingular derivative $J_n(t)$ on the target set used for inference.

(A6) *Moment and Lindeberg control.* Residual estimating contributions have finite second moments and satisfy the conditional Lindeberg condition for the displayed triangular-array or martingale approximation.

(A7) *Cutoff-process regularity.* For cutoff-indexed results, $T$ is compact or a prespecified

finite grid; the cutoff indicator class is monotone/VC; weights and nuisance scores are uniformly bounded or grid-evaluated; covariance kernels converge uniformly when interval-level weak convergence is invoked.

### Auxiliary Design Lemmas

The following elementary facts are used repeatedly below. They are stated separately to make the conditioning convention explicit.

*Lemma* (Predictable validation score). Fix a target index $t$. After conditioning on the finite archive $\mathcal{A}_n$, let $Z_i(t)$ be an integrable scalar or vector quantity that is $\mathcal{G}_{i-1}^{+}$-measurable. Suppose the recorded design law satisfies

$$\mathbb{P}(R_i = 1 \mid \mathcal{G}_{i-1}^{+}) = \pi_i, \qquad 0 < \pi_i \leq 1.$$

Set the one-step pre-draw and post-draw fields to be

$$\mathcal{H}_{i-1} := \mathcal{G}_{i-1}^{+}, \qquad \mathcal{H}_i := \mathcal{G}_{i-1}^{+} \vee \sigma(R_i),$$

and define

$$X_i(t) := \left(\frac{R_i}{\pi_i} - 1\right) Z_i(t).$$

Then $X_i(t)$ is a martingale-difference increment from $\mathcal{H}_{i-1}$ to $\mathcal{H}_i$. Equivalently,

$$X_i(t) \in L^1(\mathcal{H}_i), \qquad \mathbb{E}\{X_i(t) \mid \mathcal{H}_{i-1}\} = 0.$$

Consequently, for any fixed finite set of indices $t_1, \ldots, t_K$, the vector $(X_i(t_1), \ldots, X_i(t_K))$ is conditionally mean zero given $\mathcal{H}_{i-1}$. This martingale-difference identity uses only integrability and the recorded probability law; the second-moment, clipping, and Lindeberg assumptions enter later covariance and normal-approximation arguments, not this centering step.

*Proof.* We verify the two defining properties of a one-step martingale difference.

*Adaptation.* The pre-draw field $\mathcal{H}_{i-1}$ contains the recorded probability $\pi_i$, the finite-archive-conditioned quantity $Z_i(t)$, and all information used before the $i$th validation draw. The post-draw field $\mathcal{H}_i = \mathcal{H}_{i-1} \vee \sigma(R_i)$ additionally reveals the validation indicator. Hence

$$X_i(t) = \left(\frac{R_i}{\pi_i} - 1\right) Z_i(t)$$

is $\mathcal{H}_i$-measurable and integrable.

*Zero pre-draw conditional mean.* Since $Z_i(t)$ and $\pi_i$ are $\mathcal{H}_{i-1}$-measurable,

$$\begin{aligned}\mathbb{E}\{X_i(t) \mid \mathcal{H}_{i-1}\} &= Z_i(t)\,\mathbb{E}\left(\frac{R_i}{\pi_i} - 1 \,\middle|\, \mathcal{H}_{i-1}\right) \\ &= Z_i(t)\left\{\frac{\mathbb{E}(R_i \mid \mathcal{H}_{i-1})}{\pi_i} - 1\right\}.\end{aligned}$$

The recorded design law gives

$$\mathbb{E}(R_i \mid \mathcal{H}_{i-1}) = \mathbb{P}(R_i = 1 \mid \mathcal{H}_{i-1}) = \pi_i,$$

and therefore

$$\mathbb{E}\{X_i(t) \mid \mathcal{H}_{i-1}\} = Z_i(t)\left(\frac{\pi_i}{\pi_i} - 1\right) = 0.$$

Thus the increment is adapted after the draw and centered before the draw, which is the martingale-difference property for the filtration $\mathcal{H}_{i-1} \subseteq \mathcal{H}_i$. The vector statement follows by applying the same calculation coordinatewise. Thus the identity itself requires integrability plus the recorded probability law. The second-moment, clipping, and Lindeberg requirements used for covariance and normal approximation are stated separately in Assumption set B.1 and the CLT results below. □

**Remark (role in the A4/LEARN design).** This lemma is the basic recorded-probability calculation. The deployable rule does not need to know $Y_i(t)$ or the residual $Z_i(t)$ when it

assigns $\pi_i$. The proof conditions on the finite archive only to evaluate the randomization error of the validation design. This is the reason the paper keeps $\mathcal{D}_{i-1}$, $\mathcal{A}_n$, and $\mathcal{G}_{i-1}^{+}$ separate.

*Lemma* (Conditional covariance and predictable quadratic variation). Under the conditions of the preceding lemma, let

$$X_i(t) = \left(\frac{R_i}{\pi_i} - 1\right) Z_i(t).$$

Then, for any two fixed indices $t, t'$,

$$\mathbb{E}\{X_i(t)X_i(t') \mid \mathcal{G}_{i-1}^{+}\} = \left(\frac{1}{\pi_i} - 1\right) Z_i(t)Z_i(t').$$

Thus, for the martingale sum $M_k(t) = \sum_{i\le k} X_i(t)$, the predictable covariation is

$$\langle M(t), M(t')\rangle_n = \sum_i \left(\frac{1}{\pi_i} - 1\right) Z_i(t)Z_i(t').$$

For conditionally independent Bernoulli validation, this is also the conditional covariance contribution of the residual sum. For dependent exact-budget or without-replacement designs, the centering identity remains valid, while the finite-design covariance should use the implemented design's conditional inclusion covariance when available.

*Proof.* Because $Z_i(t)$, $Z_i(t')$, and $\pi_i$ are $\mathcal{G}_{i-1}^{+}$-measurable,

$$\mathbb{E}\{X_i(t)X_i(t') \mid \mathcal{G}_{i-1}^{+}\} = Z_i(t)Z_i(t')\,\mathbb{E}\left[\left(\frac{R_i}{\pi_i} - 1\right)^2 \middle| \mathcal{G}_{i-1}^{+}\right].$$

Since $R_i$ is an indicator and $\mathbb{E}(R_i \mid \mathcal{G}_{i-1}^{+}) = \pi_i$,

$$\mathbb{E}\left[\left(\frac{R_i}{\pi_i} - 1\right)^2 \middle| \mathcal{G}_{i-1}^{+}\right] = \frac{1}{\pi_i^2}\mathbb{E}(R_i \mid \mathcal{G}_{i-1}^{+}) - 1 = \frac{1}{\pi_i} - 1.$$

Summing these conditional second moments gives the predictable covariation. For independent Bernoulli validation, distinct increments are conditionally uncorrelated, so the same sum is the conditional covariance of the total residual process. Under dependent fixed-size designs,

cross-inclusion terms need not vanish; the exact covariance is obtained from the design's conditional inclusion covariances. □

**Remark (role in the A4/LEARN design intervals).** This lemma is the variance bridge from recorded probabilities to the reported finite-archive design intervals. It also explains the table convention: Bernoulli emulations use the displayed $(\pi_i^{-1} - 1)$ variance term, while exact-budget or fixed-count emulations require their own finite-design covariance or a clearly identified first-order approximation.

*Corollary* (Cross-fitted or pilot-learned scores). Suppose the working score $\mu_i(t)$, residual-uncertainty score $\widehat{\sigma}_i(t)$, and assigned probability $\pi_i$ are functions only of first-phase information, historical data, cross-fitting folds that exclude the unit's own final validation contribution, or pilot outcomes observed before the final-stage assignment. Assume these quantities are fixed in $\mathcal{G}_{i-1}^{+}$ before $R_i$ is drawn. Then the recorded-probability residual correction remains conditionally centered:

$$\mathbb{E}_R\left[\frac{R_i}{\pi_i}\{Y_i(t) - \mu_i(t)\}\,\middle|\,\mathcal{G}_{i-1}^{+}\right] = Y_i(t) - \mu_i(t).$$

Consequently, the validity of the residual correction does not require $\mu_i(t)$ or $\widehat{\sigma}_i(t)$ to be correctly specified. Misspecification or pilot noise affects efficiency through residual size and assigned probabilities, not the recorded-probability centering.

*Proof.* Conditional on $\mathcal{G}_{i-1}^{+}$, the quantities $\mu_i(t)$, $\widehat{\sigma}_i(t)$, and $\pi_i$ are fixed, while the finite archive $\mathcal{A}_n$ fixes $Y_i(t)$ for the design-based proof. Therefore

$$\mathbb{E}_R\left[\frac{R_i}{\pi_i}\{Y_i(t) - \mu_i(t)\}\,\middle|\,\mathcal{G}_{i-1}^{+}\right] = \{Y_i(t) - \mu_i(t)\}\frac{\mathbb{E}_R(R_i \mid \mathcal{G}_{i-1}^{+})}{\pi_i} = Y_i(t) - \mu_i(t).$$

No step uses $\mu_i(t) = \mathbb{E}\{Y_i(t) \mid W_i\}$, nor does it require the pilot score to estimate the residual uncertainty consistently. Those properties change the magnitude of the residual contribution and hence efficiency, but not the centering identity. □

**Remark (role in pilot and cross-fitted workflows).** The two-wave A4/LEARN design uses a random PET pilot to learn residual uncertainty and then records final-stage probabilities before final PET outcomes enter the estimator. Cross-fitting plays the same role for nuisance scores: it makes the score fixed for the unit whose final residual is being corrected. The result is therefore a design-validity statement, not a claim that the surrogate model is the true conditional PET risk.

## B.2 Recorded-Probability Z-Estimation

**Theorem B.1** (Recorded-probability residual correction and Z-estimation)**.** *Conditional on the first-phase archive, pilot data, and recorded validation probabilities, suppose (A1)–(A6) hold. Let* $\widehat{\theta}$ *solve* $\widehat{U}_n(\widehat{\theta}) = 0$*, and let* $\theta_n$ *solve* $U_n(\theta_n) = 0$*. Then*

$$\mathbb{E}_R\{\widehat{U}_n(\theta_n) \mid \mathcal{A}_n, \{\pi_i\}\} = U_n(\theta_n) = 0.$$

*For a direct linear target indexed by* $t$*, the recorded-probability residual correction*

$$\widehat{\theta}_n(t) = \sum_i a_i(t)\mu_i(t) + \sum_i \frac{R_i}{\pi_i} a_i(t)\{Y_i(t) - \mu_i(t)\}$$

*satisfies* $\mathbb{E}_R\{\widehat{\theta}_n(t) \mid \mathcal{A}_n, \mu, \{\pi_i\}\} = \theta_n(t)$*. Under conditionally independent Bernoulli validation, its residual process*

$$M_n(t) = \sum_i a_i(t) \left(\frac{R_i}{\pi_i} - 1\right) \{Y_i(t) - \mu_i(t)\}$$

*has conditional covariance*

$$V_R(t, t') = \sum_i \left(\frac{1}{\pi_i} - 1\right) a_i(t) a_i(t') \{Y_i(t) - \mu_i(t)\}\{Y_i(t') - \mu_i(t')\}.$$

*If* $J_n = \partial_\theta U_n(\theta_n)$ *is nonsingular and the usual Z-estimation remainder is negligible at the*

*validation scale, then*

$$\widehat{\theta} - \theta_n = -J_n^{-1}\widehat{U}_n(\theta_n) + r_n,$$

*where* $r_n = o_p(m^{-1/2})$ *under scarce-validation asymptotics and* $m = \sum_i \pi_i$*.*

*Proof. Centering.* Conditional on $\mathcal{A}_n$, the residual terms $\psi_i(\theta) - m_i(\theta)$ are fixed. For each unit and fixed $\theta$,

$$\mathbb{E}_R\left[m_i(\theta) + \frac{R_i}{\pi_i}\{\psi_i(\theta) - m_i(\theta)\}\,\middle|\,\mathcal{G}_{i-1}^{+}\right] = \psi_i(\theta),$$

because $\pi_i$ is recorded before $R_i$ and $\mathbb{E}_R(R_i \mid \mathcal{G}_{i-1}^{+}) = \pi_i$. The same identity gives the direct linear correction by replacing $\psi_i(\theta)$ with $a_i(t)Y_i(t)$ and $m_i(\theta)$ with $a_i(t)\mu_i(t)$. For conditionally independent Bernoulli validation, $\mathrm{Var}_R(R_i/\pi_i) = \pi_i^{-1} - 1$ and cross terms vanish, yielding $V_R(t, t')$. Averaging gives $\mathbb{E}_R\{\widehat{U}_n(\theta) \mid \mathcal{A}_n, \{\pi_i\}\} = U_n(\theta)$, and hence the displayed identity at $\theta = \theta_n$.

*Linearization.* Since $U_n(\theta_n) = 0$, the leading term at $\theta_n$ is the centered recorded-probability residual average. Taylor expansion around $\theta_n$ gives

$$0 = \widehat{U}_n(\theta_n) + \partial_\theta \widehat{U}_n(\widetilde{\theta})(\widehat{\theta} - \theta_n).$$

Smoothness, nuisance control, and the stated splitting conditions imply $\partial_\theta \widehat{U}_n(\widetilde{\theta}) = J_n + o_p(1)$ at the validation scale. Inverting the derivative gives

$$\widehat{\theta} - \theta_n = -J_n^{-1}\widehat{U}_n(\theta_n) + r_n,$$

with $r_n = o_p(m^{-1/2})$. □

## B.3 Scarce-Validation Asymptotics and Variance

Let $m_n = \sum_i \pi_i$, with $m_n \to \infty$ and, for scarce-validation framing, $m_n/n \to 0$.

Define the residual estimating contribution

$$e_i(\theta_n) = \psi(Y_i, Z_i; \theta_n) - m_i(\theta_n).$$

For independent Bernoulli validation conditional on the recorded probabilities,

$$\mathrm{Var}_R\{\widehat{U}_n(\theta_n) \mid \mathcal{A}_n, \{\pi_i\}\} = n^{-2} \sum_i (\pi_i^{-1} - 1) e_i e_i^\top.$$

Thus

$$\mathrm{Var}_R(\widehat{\theta} \mid \mathcal{A}_n, \{\pi_i\}) = J_n^{-1} \left[ n^{-2} \sum_i (\pi_i^{-1} - 1) e_i e_i^\top \right] J_n^{-\top} + \text{lower-order terms}.$$

**Theorem B.2** (Scarce-validation normal approximation)**.** *Under (A1)–(A6), a conditional Lindeberg condition, and* $m_n \to \infty$*, any fixed contrast* $c^\top(\widehat{\theta} - \theta_n)$ *satisfies*

$$\frac{c^\top(\widehat{\theta} - \theta_n)}{\{c^\top V_n c\}^{1/2}} \Rightarrow N(0, 1),$$

*conditionally in probability, where*

$$V_n = J_n^{-1} \left[ n^{-2} \sum_i (\pi_i^{-1} - 1) e_i e_i^\top \right] J_n^{-\top}.$$

*Proof. Array representation.* By Theorem B.1,

$$-c^\top J_n^{-1} \widehat{U}_n(\theta_n) = -\sum_i \frac{R_i - \pi_i}{\pi_i} b_i, \qquad b_i = n^{-1} c^\top J_n^{-1} e_i(\theta_n).$$

Conditional on $\mathcal{A}_n$ and the recorded probabilities, the summands form a mean-zero triangular array. Its conditional variance is

$$\sum_i (\pi_i^{-1} - 1) b_i^2 = c^\top V_n c,$$

where $V_n$ is the displayed finite-archive design covariance.

*Limit.* Positivity and the conditional Lindeberg condition give the Lindeberg-Feller central

limit theorem for the standardized array (van der Vaart, 1998). Conditional Slutsky, together with the $o_p(m_n^{-1/2})$ remainder from Theorem B.1, transfers the limit from the estimating-equation term to $c^\top(\widehat{\theta} - \theta_n)$. The same conditions give ratio consistency of the plug-in design-variance estimator. □

## B.4 Adaptive Designs and Martingale CLT

Pilot-based and sequential validation designs assign probabilities after observing earlier validation outcomes. The deployable condition is

$$\mathbb{P}(R_i = 1 \mid \mathcal{D}_{i-1}) = \pi_i, \qquad \pi_i \text{ is } \mathcal{D}_{i-1}\text{-measurable},$$

with $\pi_i$ recorded before the $i$th validation indicator is drawn. The finite-archive residual $e_i(\theta_n)$ is not required to be known to the deployable design. It is fixed only after conditioning on $\mathcal{A}_n$, which is why the proof uses $\mathcal{G}_i$ and the predictable pre-draw field $\mathcal{G}_{i-1}^+ = \mathcal{G}_{i-1} \vee \sigma(\pi_i)$.

Define

$$D_i(\theta_n) = n^{-1}\left(\frac{R_i}{\pi_i} - 1\right) e_i(\theta_n).$$

Then

$$\mathbb{E}\{D_i(\theta_n) \mid \mathcal{G}_{i-1}^+\} = 0.$$

The array is the adaptive-design analogue of the independent Bernoulli residual sum. Pilot-score misspecification can inflate the predictable quadratic variation and reduce efficiency, but it does not change the martingale centering when probabilities are predictable, recorded, and positive (Brown, 1971).

**Theorem B.3** (Adaptive recorded-probability CLT)**.** *Suppose the adaptive design satisfies positivity, predictable recorded probabilities, conditional moment bounds, and a martingale Lindeberg condition. Then the asymptotic linear representation in Theorem B.1 holds with*

$$\widehat{U}_n(\theta_n) - U_n(\theta_n) = \sum_i D_i(\theta_n),$$

*and fixed contrasts are asymptotically normal with predictable quadratic variation*

$$\langle V \rangle_n = n^{-2} \sum_i (\pi_i^{-1} - 1) e_i e_i^\top.$$

*Proof. Martingale array.* Conditional on $\mathcal{A}_n$, $e_i(\theta_n)$ is fixed. Since $\pi_i$ is known in the pre-draw field $\mathcal{G}_{i-1}^+$,

$$\mathbb{E}(R_i/\pi_i - 1 \mid \mathcal{G}_{i-1}^+) = 0,$$

so $D_i(\theta_n)$ is a martingale difference for the recorded-probability design filtration. Its predictable quadratic variation is

$$\sum_i \mathbb{E}(D_i D_i^\top \mid \mathcal{G}_{i-1}^+) = n^{-2} \sum_i (\pi_i^{-1} - 1) e_i e_i^\top,$$

the analogue of the fixed-design covariance in B.3.

*Limit and transfer.* Positivity and the martingale Lindeberg condition give the martingale central limit theorem for fixed contrasts with predictable quadratic variation $\langle V \rangle_n$ (Brown, 1971). The Z-estimation expansion in Theorem B.1 then transfers the limit to the estimator. In the two-wave A4/LEARN emulation, the random pilot affects the final-stage probabilities through $\mathcal{D}_{i-1}$; after those probabilities are assigned and recorded, the final residual correction is centered under $\mathcal{G}_i$. □

## B.4a Superpopulation Efficient-Information Bridge

The main manuscript and the reported intervals are finite-archive and design-based. This subsection records the corresponding superpopulation interpretation for readers who want the classical semiparametric efficient-information calculation (Bickel et al., 1993; van der Vaart, 1998; Tsiatis, 2006). It is a bridge to the missing-data efficiency literature, not an

additional empirical claim for A4/LEARN.

Let $X = (W, Y)$ denote full data and let $O = (W, R, RY)$ denote observed data under a known recorded validation law

$$\mathbb{P}(R = 1 \mid W, Y) = \mathbb{P}(R = 1 \mid W) = \pi(W), \qquad 0 < \pi_{\min} \le \pi(W) \le 1.$$

For a smooth superpopulation target $\theta_0$ solving

$$P\{\psi(Y, W; \theta_0)\} = 0, \qquad J = P\{\partial_\theta \psi(Y, W; \theta_0)\},$$

write

$$\eta(W; \theta_0) = \mathbb{E}\{\psi(Y, W; \theta_0) \mid W\}.$$

*Proposition* (Semiparametric efficient-information bridge). In the nonparametric observed-data model with known validation probability $\pi(W)$, the efficient influence function for $\theta_0$ is

$$\phi_{\text{eff}}(O) = -J^{-1}\left[\eta(W; \theta_0) + \frac{R}{\pi(W)}\{\psi(Y, W; \theta_0) - \eta(W; \theta_0)\}\right].$$

Thus the observed-data efficiency bound is

$$V_{\text{eff}} = J^{-1}\operatorname{Var}\left[\eta(W; \theta_0) + \frac{R}{\pi(W)}\{\psi(Y, W; \theta_0) - \eta(W; \theta_0)\}\right] J^{-\top}.$$

For the scalar linear target $X_t = a_t(W)Y(t)$, with $\theta_0(t) = \mathbb{E}X_t$ and $\eta_t(W) = \mathbb{E}(X_t \mid W)$, this reduces to

$$\phi_{\text{eff},t}(O) = \eta_t(W) - \theta_0(t) + \frac{R}{\pi(W)}\{X_t - \eta_t(W)\},$$

and

$$V_{\text{eff}}(t, t') = \operatorname{Cov}\{\eta_t(W), \eta_{t'}(W)\} + \mathbb{E}\left[\frac{1}{\pi(W)}\operatorname{Cov}\{X_t, X_{t'} \mid W\}\right].$$

If a recorded-probability estimator uses an arbitrary first-phase control variate $m(W; \theta_0)$,

its influence function is the same display with $\eta$ replaced by $m$. Its covariance exceeds the efficient bound by

$$J^{-1}\mathbb{E}\left[\left\{\frac{1}{\pi(W)}-1\right\}\{m(W;\theta_0)-\eta(W;\theta_0)\}\{m(W;\theta_0)-\eta(W;\theta_0)\}^\top\right]J^{-\top}.$$

Consequently the recorded-probability estimator is semiparametrically efficient in this super-population MAR model when the control variate consistently estimates the conditional mean $\eta(W;\theta_0)$, with nuisance-estimation remainders negligible at the validation scale.

*Proof.* The known-probability missing-data tangent-space calculation maps a full-data influence contribution $h(X)$ with mean zero to the observed-data contribution (Bickel et al., 1993; Tsiatis, 2006)

$$\mathbb{E}\{h(X) \mid W\} + \frac{R}{\pi(W)}\left[h(X) - \mathbb{E}\{h(X) \mid W\}\right].$$

Applying this standard coarsening-at-random representation to $h(X) = -J^{-1}\psi(Y, W; \theta_0)$ gives the displayed efficient influence function.

For the variance comparison, set $\delta(W) = m(W;\theta_0) - \eta(W;\theta_0)$ and $\varepsilon = \psi(Y, W;\theta_0) - \eta(W;\theta_0)$. Under the recorded validation law, $R$ is conditionally independent of $\varepsilon$ given $W$, $\mathbb{E}(\varepsilon \mid W) = 0$, and

$$m + \frac{R}{\pi}(\psi - m) = \eta + \frac{R}{\pi}\varepsilon + \left(1 - \frac{R}{\pi}\right)\delta.$$

The last term has conditional mean zero and is conditionally uncorrelated with $\eta + R\varepsilon/\pi$. Its conditional variance is $(\pi^{-1} - 1)\delta\delta^\top$, which gives the positive-semidefinite excess covariance above. Hence the conditional-mean control variate is the unique variance minimizer in this augmented inverse-probability class, up to $L_2(P)$ equivalence on the validated support. □

For design allocation, the efficient-information display leaves only the residual term depending on the validation probability. For the scalar linear target,

$$V_{\text{eff}}(t,t) = \operatorname{Var}\{\eta_t(W)\} + \mathbb{E}\left[\frac{a_t(W)^2 \operatorname{Var}\{Y(t) \mid W\}}{\pi(W)}\right].$$

The first term is not affected by the validation design. Under an expected cost constraint, the first-order efficient-information allocation therefore samples in proportion to

$$\frac{|a_t(W)| \ \mathrm{Var}\{Y(t) \mid W\}^{1/2}}{\sqrt{c(W)}},$$

with clipping under positivity bounds. The finite-archive rule in the manuscript is the design-based analogue of this calculation, replacing conditional residual standard deviation by the recorded residual-uncertainty score $\sigma_i(t)$.

## B.5 Cutoff-Indexed Weak Convergence and Multiplier Bands

Let $C_i$ be a continuous or ordered protocol measurement such as Centiloid, and let $T = [c_L, c_U]$ be a prespecified cutoff interval. For $c \in T$, define

$$Y_i(c) = 1(C_i \geq c), \qquad \theta_n(c) = \sum_i a_i(c) Y_i(c), \qquad e_i(c) = Y_i(c) - \mu_i(c).$$

The augmented cutoff-indexed estimator is

$$\widehat{\theta}(c) = \sum_i a_i(c)\mu_i(c) + \sum_i \frac{R_i}{\pi_i} a_i(c)\{Y_i(c) - \mu_i(c)\}.$$

For direct linear contrasts, the validation residual process is

$$G_n(c) = \sqrt{m_n}\{\widehat{\theta}(c) - \theta_n(c)\} = \sqrt{m_n} \sum_i \left(\frac{R_i}{\pi_i} - 1\right) a_i(c) e_i(c),$$

up to nuisance-estimation remainders. For a finite cutoff set such as the reported 20/24/30 grid, $G_n$ is a finite-dimensional vector and joint normality follows from Theorem B.2. For a compact cutoff interval, the same centered design error is treated as a random element of $\ell^\infty(T)$. The monotone cutoff class, positivity, boundedness, and a uniform Lindeberg condition support tightness and convergence to a mean-zero Gaussian process with covariance kernel $K(c, c')$; this is the supplement-facing empirical-process formulation used for simultaneous

cutoff bands (van der Vaart and Wellner, 1996). For an interval, use the empirical-process conditions below.

(P1) $T$ is compact and $\{1(C_i \geq c) : c \in T\}$ is treated as a VC/monotone indicator class.

(P2) $a_i(c)$ and $\mu_i(c)$ are uniformly bounded and sufficiently regular over $c$, or are evaluated on a prespecified fine grid.

(P3) Learned $\mu_i(c)$ is fixed conditional on the design history, or is learned by pilot/main splitting or cross-fitting.

(P4)

$$K_n(c, c') = m_n \sum_i a_i(c) a_i(c') (\pi_i^{-1} - 1) e_i(c) e_i(c')$$

converges uniformly to a continuous covariance kernel $K(c, c')$.

(P5) A uniform Lindeberg condition holds, with validation probabilities bounded away from zero.

(P6) Either Centiloid has no large atoms on $T$, or inference is reported on a finite cutoff grid.

*Proposition* (Finite-grid cutoff max statistic). Let $T_K = \{c_1, \dots, c_K\}$ be a fixed prespecified cutoff grid. For each $c_k$, let

$$G_n(c_k) = \sqrt{m_n} \sum_i \left( \frac{R_i}{\pi_i} - 1 \right) a_i(c_k) e_i(c_k),$$

with finite-archive covariance matrix $K_n$ whose $(k, \ell)$ entry is

$$K_n(c_k, c_\ell) = m_n \sum_i (\pi_i^{-1} - 1) a_i(c_k) a_i(c_\ell) e_i(c_k) e_i(c_\ell).$$

Under the fixed-grid version of the conditional Lindeberg condition and $K_n \to K$, where $K$ is positive definite on the reported grid,

$$\{G_n(c_1), \dots, G_n(c_K)\} \Rightarrow N_K(0, K)$$

conditionally in probability. Hence the design-based max statistic

$$T_n = \max_{1\le k\le K} \frac{|G_n(c_k)|}{K_n(c_k,c_k)^{1/2}}$$

converges to

$$T = \max_{1\le k\le K} \frac{|Z_k|}{K(c_k,c_k)^{1/2}}, \qquad Z \sim N_K(0,K).$$

If $\widehat{K}_n$ is conditionally consistent for $K_n$, the Gaussian multiplier statistic computed with covariance $\widehat{K}_n$ gives asymptotic critical values for simultaneous inference on the fixed grid, provided the limit distribution is continuous at the chosen quantile.

*Proof.* For any fixed vector $u \in \mathbb{R}^K$,

$$\sum_{k=1}^{K} u_k G_n(c_k) = \sqrt{m_n} \sum_i \left(\frac{R_i}{\pi_i} - 1\right) \sum_{k=1}^{K} u_k a_i(c_k) e_i(c_k).$$

The inner finite sum is fixed after conditioning on $\mathcal{A}_n$ and the pre-draw design fields. The conditional Lindeberg condition therefore gives the one-dimensional central limit theorem for every $u$. Cramer-Wold yields the multivariate normal limit with covariance $K$ (van der Vaart, 1998). The map

$$x \mapsto \max_{1\le k\le K} |x_k|/K(c_k,c_k)^{1/2}$$

is continuous when the diagonal entries are positive, so the continuous mapping theorem gives the limit of $T_n$. If $\widehat{K}_n$ is consistent, then the conditional Gaussian multiplier vector with covariance $\widehat{K}_n$ converges to the same $N_K(0,K)$ law. Continuity of the limiting max distribution at the target quantile gives consistency of the multiplier critical value. □

**Remark (role in cutoff sensitivity).** The reported CL20/CL24/CL30 cutoff summaries are finite-dimensional. They can be read through this proposition without invoking the continuous-index $\ell^\infty(T)$ machinery. The continuous process result remains useful for the supplement-facing multiplier band over a cutoff interval, but the finite-grid claim already

covers the displayed cutoff table and pointwise grid summaries.

**Proposition B.4** (Cutoff-indexed recorded-probability process)**.** *Under (A1)–(A7) and (P1)–(P6), conditional on* $\mathcal{A}_n$ *and recorded validation probabilities,*

$$G_n(\cdot) \Rightarrow G(\cdot)$$

*in* $\ell^\infty(T)$*, where* $G(\cdot)$ *is a mean-zero Gaussian process with covariance kernel* $K(c, c')$*. For a Z-estimation target indexed by* $c$*, the same conclusion follows after uniform linearization, provided the Jacobian matrices are uniformly nonsingular and the nuisance/design remainders are uniformly negligible.*

*Proof.* Fix $c_1, \ldots, c_k \in T$. Conditional on $\mathcal{A}_n$ and on the recorded probabilities, the vector $(G_n(c_1), \ldots, G_n(c_k))$ is a finite-dimensional recorded-probability residual array. Each summand is centered by the same calculation as in Theorem B.1, and its covariance is the finite-dimensional restriction of $K_n$. The multivariate Lindeberg-Feller theorem, or the martingale version when probabilities are assigned adaptively, gives Gaussian finite-dimensional limits (van der Vaart, 1998; Brown, 1971).

It remains to control the index $c$. The class $c \mapsto 1(C_i \geq c)$ is monotone and hence VC; multiplying by uniformly bounded target weights and fixed nuisance functions preserves polynomial entropy for the residual class. Positivity controls the inverse-probability multipliers uniformly, and the uniform Lindeberg condition rules out a dominating unit over the interval. Standard empirical-process maximal inequalities for such bounded VC classes give asymptotic equicontinuity in $\ell^\infty(T)$ (van der Vaart and Wellner, 1996). Uniform convergence of $K_n(c, c')$ identifies the covariance kernel of the limit process. Thus $G_n \Rightarrow G$ in $\ell^\infty(T)$. If Centiloid has ties or large atoms on $T$, the finite-grid conclusion follows from the finite-dimensional argument without invoking interval-level tightness. For a Z-estimation target, apply the uniform version of the Taylor expansion used in Theorem B.1; uniform nonsingularity of the Jacobian and negligible nuisance/design remainders transfer the process limit from the

estimating equation to the estimator. □

The covariance kernel can be estimated by

$$\widehat{K}(c,c') = m_n \sum_i \frac{R_i(1-\pi_i)}{\pi_i^2} a_i(c)a_i(c')\widehat{e}_i(c)\widehat{e}_i(c'),$$

or conservatively by omitting the $(1-\pi_i)$ factor. A simultaneous band can be built with multiplier draws

$$\widehat{G}^*(c) = \sqrt{m_n} \sum_i \xi_i \frac{\{R_i(1-\pi_i)\}^{1/2}}{\pi_i} a_i(c)\widehat{e}_i(c),$$

using the conditional quantile of $\sup_c |\widehat{G}^*(c)|/\widehat{s}(c)$, where $\widehat{s}(c)^2 = \widehat{K}(c,c)$.

### Connection to the Cutoff-Sensitivity Figures

For the cutoff-indexed target, the recorded-probability estimator applies pointwise in $c$. On the reported 20/24/30 grid, joint normality follows from the multivariate version of Theorem B.2. Over an interval, the monotone threshold class supports weak convergence in $\ell^\infty(T)$ under the stated uniform Lindeberg and nuisance-control conditions. The multiplier-band calculation in Supplement Figure D.1 is the process analogue of the pointwise cutoff summaries in Supplement D.2.

## B.6 First-Order Allocation Programs for Recorded Validation Values

**Proposition B.5** (First-order allocation programs for recorded validation values)**.** *The scalar, cost-weighted, and integrated cutoff-range allocation problems below are first-order Neyman/KKT calculations for recorded-probability validation values, with roots in classical allocation, optimal two-phase design, and optimal subsampling (Neyman, 1934; Cochran, 1977; Särndal et al., 1992; Gilbert et al., 2014; Fithian and Hastie, 2014; Wang et al., 2018, 2019; Imberg et al., 2025). The programs are convex in $\pi$ on $\pi_i > 0$; their interior KKT solutions sample in proportion to the square root of the target-specific variance contribution,*

*with cost adjustment and probability clipping when bounds are imposed. Oracle rows in the empirical tables are diagnostic lower bounds based on full-protocol residual information.*

For a scalar target, the first-order validation objective is

$$\min_{\pi} \quad \sum_i \frac{a_i^2 \sigma_i^2}{\pi_i}$$

$$\text{subject to} \quad \sum_i \pi_i = m, \qquad \pi_{\min} \leq \pi_i \leq 1.$$

Ignoring truncation,

$$\pi_i^\star \propto |a_i| \sigma_i.$$

With unit-specific costs $c_i$, solve

$$\min_{\pi} \quad \sum_i \frac{a_i^2 \sigma_i^2}{\pi_i}$$

$$\text{subject to} \quad \sum_i c_i \pi_i = B, \qquad \pi_{\min} \leq \pi_i \leq 1.$$

The interior solution is

$$\pi_i^\star \propto \frac{|a_i| \sigma_i}{\sqrt{c_i}}.$$

For a cutoff-indexed target over a finite grid $T$, an integrated objective is

$$\min_{\pi} \quad \sum_i \frac{\sum_{c \in T} \omega_c a_i(c)^2 \sigma_i(c)^2}{\pi_i},$$

with interior solution

$$\pi_i^\star \propto \left\{ \sum_{c \in T} \omega_c a_i(c)^2 \sigma_i(c)^2 \right\}^{1/2}.$$

With unit-specific costs, divide the score by $\sqrt{c_i}$. With lower and upper probability bounds, the KKT solution is the clipped version:

$$\pi_i^\star = \min\left[1,\ \max\left\{\pi_{\min},\ \kappa\,\frac{\left(\sum_{c\in T}\omega_c a_i(c)^2\sigma_i(c)^2\right)^{1/2}}{\sqrt{c_i}}\right\}\right].$$

For worst-case cutoff performance on a finite grid, solve the convex epigraph problem

$$\begin{aligned}
\min_{\pi,z} \quad & z \\
\text{subject to} \quad & \sum_i \frac{a_i(c)^2\sigma_i(c)^2}{\pi_i} \le z, \qquad c \in T, \\
& \sum_i c_i\pi_i \le B, \qquad \pi_{\min} \le \pi_i \le 1.
\end{aligned}$$

At an interior optimum, KKT conditions give the same form as the integrated design with weights on the active cutoff constraints.

*Proof.* Let $q_i \ge 0$ denote the target-specific variance contribution: $q_i = a_i^2\sigma_i^2$ for a scalar contrast, and $q_i = \sum_{c\in T}\omega_c a_i(c)^2\sigma_i(c)^2$ for the integrated cutoff objective. The map $\pi_i \mapsto q_i/\pi_i$ is convex on $\pi_i > 0$, so the displayed programs are convex over the feasible probability region.

For the interior cost-weighted problem, the Lagrangian is

$$L(\pi,\lambda) = \sum_i \frac{q_i}{\pi_i} + \lambda\left(\sum_i c_i\pi_i - B\right).$$

The first-order condition

$$-\frac{q_i}{\pi_i^2} + \lambda c_i = 0$$

gives $\pi_i \propto \sqrt{q_i/c_i}$. Substituting the scalar or integrated $q_i$ gives the displayed allocation values. Lower and upper probability bounds enter through complementary slackness and yield the clipped form. The maximin finite-grid problem is a convex epigraph problem; KKT multipliers on the active cutoff constraints create nonnegative active-cutoff weights, producing the same square-root allocation form with an active weighted average of cutoff-specific $q_i(c)$. □

*Lemma* (Recorded-probability comparator designs). Let $s_i \ge 0$ be any design score fixed before the $i$th validation draw, such as a constant score for random validation, a stratum-balancing

score, a generic uncertainty score, or a target-specific value score. Let the implemented rule map the scores, costs, strata, and budget constraints to positive recorded probabilities $\pi_i$, possibly through clipping:

$$\pi_i(\lambda) = \min\{1, \max(\pi_{\min}, \lambda s_i/\sqrt{c_i})\}.$$

If the implemented design satisfies

$$\mathbb{P}(R_i = 1 \mid \mathcal{G}_{i-1}^{+}) = \pi_i,$$

then the recorded-probability residual correction is conditionally centered for random, balanced, uncertainty-based, target-specific, pilot-learned, and clipped probability rows under the same argument. Under conditionally independent Bernoulli validation, the covariance contribution is

$$\sum_i \left(\frac{1}{\pi_i} - 1\right) a_i(t)a_i(t')e_i(t)e_i(t').$$

For exact-budget, stratified fixed-count, PPS-like, or without-replacement implementations, the centering still follows from the recorded first-order inclusion probabilities, while the exact finite-design covariance uses the corresponding conditional pairwise inclusion probabilities.

*Proof.* The clipping and budget-normalization step produces probabilities that are fixed before the validation indicator is drawn. Once those probabilities are recorded, the score that generated them is irrelevant to the centering calculation. Applying the predictable validation score lemma with

$$Z_i(t) = a_i(t)e_i(t)$$

gives

$$\mathbb{E}_R\left[\left(\frac{R_i}{\pi_i} - 1\right) a_i(t)e_i(t) \,\middle|\, \mathcal{G}_{i-1}^{+}\right] = 0.$$

Summing over $i$ gives conditional centering for any comparator row satisfying the recorded

design law and positivity. If the validation indicators are conditionally independent Bernoulli, the conditional covariance follows from the conditional covariance lemma. If the design fixes counts or samples without replacement, then the same Horvitz-Thompson centering holds under the recorded first-order inclusion probabilities, but cross terms may remain. The exact variance is then

$$\sum_{i,j} a_i(t)a_j(t')e_i(t)e_j(t') \operatorname{Cov}\left(\frac{R_i}{\pi_i}, \frac{R_j}{\pi_j} \,\middle|\, \mathcal{A}_n, \{\pi_i\}\right),$$

or a stated first-order approximation when the exact pairwise inclusion covariance is not part of the implemented emulation. □

**Remark (role in comparator rows).** This lemma is why random validation, APOE4-balanced validation, generic uncertainty validation, target-specific validation, and pilot-learned validation can be displayed on the same inferential scale. They are different ways to assign recorded probabilities, not different estimators. The KKT calculations motivate one class of scores; they do not create a new universal allocation theorem or a dominance claim over balanced or active-inference alternatives.

## B.7 Uncertainty-Sampling Regret

Define first-order validation variance

$$V(\pi;\theta) = \sum_i \frac{a_i^2\sigma_i^2}{\pi_i}.$$

Let $\pi^\star$ be the target-optimal design and let $\pi^U$ be an uncertainty design with $\pi_i^U \propto \sigma_i$. Define validation regret

$$\text{Regret}(\pi;\theta) = \frac{V(\pi;\theta)}{V(\pi^\star;\theta)} - 1.$$

**Proposition B.6** (Uncertainty-sampling separation)**.** *There exists a two-stratum sequence of valid contrast-inference problems such that*

$$\text{Regret}(\pi^U;\theta) \to \infty,$$

*while target-optimal validation has bounded variance under the same expected validation budget.*

*Construction.* Take two strata, $H$ and $L$, with equal numbers of units and a common expected validation fraction $\rho$. Set

$$|a_H| = M, \quad \sigma_H = M^{-1}, \qquad |a_L| = 1, \quad \sigma_L = 1,$$

with $M \to \infty$. The target-specific variance contributions satisfy $a_H^2\sigma_H^2 = a_L^2\sigma_L^2 = 1$, so the target-optimal design validates the two strata at the same order and has $V(\pi^\star;\theta) \asymp \rho^{-1}$. Uncertainty sampling allocates in the score ratio $M^{-1} : 1$, so $\pi_H^U \asymp \rho/M$. Its $H$-stratum contribution satisfies $a_H^2\sigma_H^2/\pi_H^U \asymp M\rho^{-1}$, and therefore $V(\pi^U;\theta)/V(\pi^\star;\theta) \to \infty$. □

The divergence statement is for the unclipped first-order allocation, or for a scarce-validation sequence in which the effective lower bound does not bind the undersampled stratum. With a fixed clipping bound, the same construction gives a finite but potentially large separation up to the clipping constraint.

## B.8 Pilot-Estimated Design Robustness

Let $s_i = |a_i|\sigma_i$ be the oracle score and let $\widehat{s}_i$ be a pilot-estimated score. Suppose, after truncation,

$$K^{-1} \le \frac{\widehat{s}_i}{s_i} \le K$$

for all units with nonzero score. Then the validation variance of the pilot design is bounded by a multiplicative function of $K$ relative to the oracle design.

**Proposition B.7** (Multiplicative pilot-score robustness). *Under bounded multiplicative score error and common positivity truncation, the pilot-estimated design has*

$$\frac{V(\widehat{\pi};\theta)}{V(\pi^\star;\theta)} \le K^2$$

*up to constants induced by clipping and budget normalization.*

*Proof.* Without clipping,

$$\pi_i^\star = \frac{ms_i}{\sum_j s_j}, \qquad \widehat{\pi}_i = \frac{m\widehat{s}_i}{\sum_j \widehat{s}_j}.$$

The bound $K^{-1} \le \widehat{s}_i/s_i \le K$ implies

$$K^{-2} \le \frac{\widehat{\pi}_i}{\pi_i^\star} \le K^2.$$

Clipping changes only the constants determined by the common bounds. Thus $1/\widehat{\pi}_i \le C_K/\pi_i^\star$ uniformly over nonzero-score units, and summing $a_i^2\sigma_i^2/\widehat{\pi}_i$ gives the stated variance comparison. If $K$ is large, the design can lose efficiency, but the recorded-probability correction remains centered. □

## B.9 Variance Accounting for Reported Intervals

This section is the supplement reference for the manuscript's variance-accounting statement. Unless otherwise stated, variance statements condition on the finite A4/LEARN analysis archive through $\mathcal{A}_n$, the deployable history used to assign probabilities, the working surrogate/control-variate scores, and the recorded validation probabilities. The reported confidence intervals are finite-archive design intervals: they quantify uncertainty from the validation indicators $R_i$ under the recorded-probability residual correction.

The included component is the design variance of the inverse-probability residual correction. For independent Bernoulli validation, this is the residual variance in B.3. For exactly budgeted, clipped, PPS-like, or without-replacement emulations, the displayed variance is the

corresponding first-order recorded-probability approximation unless a script-specific output reports the exact finite-sample covariance. For pilot or adaptive designs, B.4 treats assigned probabilities as predictable and uses the probabilities recorded before outcome contribution. Repeated pilot summaries also reflect pilot-wave randomness and learned scores, whereas a realized-design interval conditions on the recorded design.

Random, uncertainty, and target-specific retrospective emulations are finite-archive policy evaluations under recorded validation probabilities and fixed budget labels. Oracle lower-bound rows use full-protocol residual information and are evaluation-only. Two-wave pilot rows use the stated pilot/final-wave budget split; final inference conditions on probabilities assigned after the pilot and recorded before final-stage outcome use. Cutoff-indexed rows report pointwise finite-archive intervals, with the simultaneous multiplier calculation shown as process evidence in the supplement.

The reported design intervals do not add separate variance components for superpopulation sampling of A4/LEARN participants, model selection for the working surrogate score, feature-engineering decisions, or uncertainty about whether each auxiliary biomarker would be available in a future prospective workflow. Cross-fitting, sample splitting, or pilot/main separation is used to avoid direct residual reuse; remaining nuisance estimation is handled through the stated remainder conditions and sensitivity analyses rather than through a full bootstrap over model construction. PET is treated as the protocol measurement defining the target, so biological measurement error beyond that protocol is not included in the validation-design variance.

For cutoff-indexed targets, pointwise intervals are finite-archive design intervals at each cutoff. The multiplier band in B.5 is simultaneous process evidence for the prespecified cutoff range.

## B.10 Product-Integral Bivariate Survival Targets

This subsection gives the finite-grid bivariate survival version of the recorded-probability argument. It is included because many protocol measurements are naturally paired event times, such as two time-to-threshold outcomes or a time-to-event and time-to-protocol-confirmation pair. The design principle is unchanged: the validation probability is recorded before the paired protocol outcome enters the estimator, and misspecification of the first-phase survival surface affects efficiency rather than design centering.

The role of this subsection is extension and positioning, not a new A4/LEARN bivariate survival analysis. The real-data A4/LEARN analyses demonstrate the recorded-probability workflow for scalar PET contrasts and for the cutoff-indexed PET positivity curve $Y_i(c) = 1(C_i \geq c)$. The same finite-grid algebra is what supports the CL20/CL24/CL30 audit and the cutoff-process multiplier band in Supplement B.5 and Supplement D.2. The present subsection records how the same design contract would be used when the protocol measurement is a paired event-time surface. No bivariate survival simulation is included in the submission supplement because the empirical paper is about A4/LEARN amyloid PET, and simulation displays are kept to mechanisms that are directly used to interpret the PET validation design.

Let $T_i = (T_{i1}, T_{i2})$ be the paired protocol time for unit $i$. For $(s, t)$ in a prespecified finite grid $\mathcal{K} = \{(s_\ell, t_\ell) : \ell = 1, \ldots, q\}$, define

$$Y_i(s,t) = 1(T_{i1} > s, T_{i2} > t), \qquad \theta_n(s,t) = \sum_i a_i(s,t) Y_i(s,t).$$

The weights $a_i(s,t)$ encode the registered surface contrast. For example, a high-risk minus low-risk contrast averaged over the grid uses

$$a_i(s_\ell, t_\ell) = q^{-1} \left\{ \frac{1(G_i = 1)}{n_1} - \frac{1(G_i = 0)}{n_0} \right\}, \qquad \ell = 1, \ldots, q,$$

where $n_g = \sum_i 1(G_i = g)$. Other choices encode a weighted surface average, a regression-slope

surface, or a finite collection of clinically selected grid points.

A first-phase or pilot-fitted bivariate survival working model supplies $\mu_i(s,t)$. One natural notation is the product-integral form associated with Dabrowska-type bivariate survival estimators (Dabrowska, 1988; Andersen et al., 1993; Hougaard, 2000):

$$\mu_i(s,t) = \prod_{u\le s}\{1-\Lambda_{10,i}(du,0)\}\prod_{v\le t}\{1-\Lambda_{01,i}(0,dv)\}\prod_{u\le s,\, v\le t}\{1-L_i(du,dv)\}.$$

Here $\Lambda_{10,i}$ and $\Lambda_{01,i}$ are working conditional marginal cumulative hazard measures, and $L_i$ is a working bivariate dependence increment. The product-integral display is a compact way to define the control variate. It can be replaced by any first-phase or pilot-learned surface fixed before the corresponding validation outcome is used; the validation proof below does not require the working surface to be correctly specified.

For $(s,t)\in\mathcal{K}$, use

$$\widehat{\theta}_n(s,t) = \sum_i a_i(s,t)\mu_i(s,t) + \sum_i \frac{R_i}{\pi_i} a_i(s,t)\{Y_i(s,t)-\mu_i(s,t)\}.$$

Stack the $q$ grid values in $\widehat{\boldsymbol{\theta}}_{\mathcal{K}}$, and define the residual-contribution vector

$$\boldsymbol{d}_i = \Big(a_i(s_\ell,t_\ell)\{Y_i(s_\ell,t_\ell)-\mu_i(s_\ell,t_\ell)\}\Big)_{\ell=1}^{q}.$$

Let $H$ be any fixed matrix that maps the grid vector to the registered summaries, such as selected grid components, an average surface contrast, or a small set of clinically weighted contrasts.

**Proposition B.8** (Finite-grid bivariate survival validation process)**.** *Condition on the finite archive, the working product-integral surfaces, the registered grid $\mathcal{K}$, the weights $a_i(s,t)$, and recorded positive probabilities $\pi_i$. If*

$$\mathbb{E}(R_i \mid \mathcal{G}_{i-1}^{+}) = \pi_i, \qquad 0<\pi_i\le 1,$$

*with $\pi_i$ recorded before the paired protocol outcome contributes to estimation, then*

$$\widehat{\boldsymbol{\theta}}_{\mathcal{K}} - \boldsymbol{\theta}_{\mathcal{K}} = \sum_i \left(\frac{R_i}{\pi_i} - 1\right) \boldsymbol{d}_i, \qquad \mathbb{E}_R(\widehat{\boldsymbol{\theta}}_{\mathcal{K}} \mid \mathcal{A}_n, \pi) = \boldsymbol{\theta}_{\mathcal{K}},$$

*with the first display interpreted as a martingale sum for adaptive validation. Under conditionally independent Bernoulli validation,*

$$V_{\mathcal{K}}(\pi) = \mathrm{Var}_R(\widehat{\boldsymbol{\theta}}_{\mathcal{K}} \mid \mathcal{A}_n, \pi) = \sum_i \left(\frac{1}{\pi_i} - 1\right) \boldsymbol{d}_i \boldsymbol{d}_i^{\top}.$$

*For adaptive validation, the same expression is the predictable quadratic-variation analogue when the draws are conditionally Bernoulli at their recorded probabilities.*

*If $V_{\mathcal{K}}(\pi)$ is nondegenerate on the registered contrast space and*

$$\max_i \left\| \{H V_{\mathcal{K}}(\pi) H^{\top}\}^{-1/2} \left(\frac{1}{\pi_i} - 1\right)^{1/2} H \boldsymbol{d}_i \right\| \longrightarrow 0,$$

*then*

$$\{H V_{\mathcal{K}}(\pi) H^{\top}\}^{-1/2} H(\widehat{\boldsymbol{\theta}}_{\mathcal{K}} - \boldsymbol{\theta}_{\mathcal{K}}) \Rightarrow N(0, I).$$

*The covariance can be estimated from validated residuals by*

$$\widehat{V}_{\mathcal{K}} = \sum_i \frac{R_i(1 - \pi_i)}{\pi_i^2} \boldsymbol{d}_i \boldsymbol{d}_i^{\top},$$

*and a corresponding Gaussian multiplier draw is*

$$\widehat{G}^*_{\mathcal{K}} = \sum_i \xi_i \frac{R_i(1 - \pi_i)^{1/2}}{\pi_i} \boldsymbol{d}_i,$$

*with independent standard normal multipliers $\xi_i$. Let $c^*_{1-\alpha}$ be the conditional quantile of*

$$\max_r \left| \{H \widehat{G}^*_{\mathcal{K}}\}_r \right| / \widehat{\mathrm{se}}_r,$$

*where* $\widehat{\mathrm{se}}_r^2$ *is the* $r$*th diagonal element of* $H\widehat{V}_{\mathcal{K}}H^\top$*. Then*

$$\{H\widehat{\boldsymbol{\theta}}_{\mathcal{K}}\}_r \pm c^*_{1-\alpha}\widehat{\mathrm{se}}_r$$

*gives the finite-grid simultaneous recorded-probability band for the registered summaries.*

*Proof.* After conditioning on $\mathcal{A}_n$, $Y_i(s,t)$, $\mu_i(s,t)$, $a_i(s,t)$, $H$, and $\pi_i$ are fixed for every grid point. The recorded-probability condition gives

$$\mathbb{E}\{(R_i/\pi_i - 1)\boldsymbol{d}_i \mid \mathcal{G}^+_{i-1}\} = 0,$$

so the stacked residual correction is centered componentwise and is a martingale transform in the adaptive case. Conditional Bernoulli independence gives the covariance sum; otherwise the displayed sum is the predictable quadratic variation when the conditional variance is $\pi_i(1-\pi_i)$. Applying Cramer-Wold to $H\boldsymbol{d}_i$, the stated maximum-leverage condition is the finite-dimensional Lindeberg condition and yields the normal approximation (van der Vaart, 1998; Brown, 1971). The plug-in covariance is the usual recorded-probability residual covariance estimate because $\mathbb{E}_R\{R_i(1-\pi_i)\boldsymbol{d}_i\boldsymbol{d}_i^\top/\pi_i^2\} = (1/\pi_i - 1)\boldsymbol{d}_i\boldsymbol{d}_i^\top$. Conditional on the validated residuals, the multiplier draw has covariance $\widehat{V}_{\mathcal{K}}$, and the max statistic gives the stated finite-grid simultaneous band (van der Vaart and Wellner, 1996). □

For design allocation, choose $H$ before validation. The first-order trace objective is

$$\mathrm{tr}\{HV_{\mathcal{K}}(\pi)H^\top\} = \sum_i \left(\frac{1}{\pi_i} - 1\right)\|H\boldsymbol{d}_i\|^2.$$

Dropping the constant $-\sum_i \|H\boldsymbol{d}_i\|^2$, the expected-budget program minimizes $\sum_i \|H\boldsymbol{d}_i\|^2/\pi_i$ subject to $\sum_i \pi_i = m$ and positivity bounds. The unclipped interior allocation is therefore proportional to $\|H\boldsymbol{d}_i\|$. In design-anticipated form this becomes

$$s_i = \left[\mathrm{tr}\left\{HA_i\Gamma_iA_i^\top H^\top\right\}\right]^{1/2},$$

where $A_i$ is the diagonal matrix of grid weights $a_i(s_\ell, t_\ell)$ and $\Gamma_i$ is the working residual covariance matrix of $\{Y_i(s_\ell, t_\ell) - \mu_i(s_\ell, t_\ell)\}_{\ell=1}^q$ given first-phase information. This is the bivariate-survival analogue of the scalar rule $|a_i|\sigma_i$: a unit is valuable when it has influence for the registered surface summary and residual protocol uncertainty after the working survival surface.

This extension is anchored by the real-data A4/LEARN cutoff results, which are the one-index version of the same recorded-probability calculation. There the correction is applied to $Y_i(c) = 1(C_i \geq c)$, and Supplement D.2 reports stable APOE4 PET contrasts at CL20, CL24, and CL30 with integrated cutoff-range CI-width ratios 0.861, 0.859, and 0.918 relative to random validation at budget 200. The paired-time product-integral notation is therefore included as a mathematical extension of the recorded-probability contract, not as a second empirical or simulation contribution. A submission-facing simulation should either reproduce the scalar/cutoff PET validation logic used in the paper or be replaced by real paired-time data; this supplement takes the conservative route and keeps the paired-time material theoretical.

## B.11 Pairwise or Nonlinear Protocol Targets

The A4/LEARN empirical targets analyzed here are linear finite-archive contrasts and cutoff-indexed linear contrasts; the product-integral bivariate survival material is a formal finite-grid extension. Pairwise agreement, AUC-type, rank, or kernel targets would require a different target definition, pairwise validation probabilities, kernel entropy conditions, and control of degenerate U-process remainders (Arcones and Giné, 1993). The present supplement therefore stops before these nonlinear targets.

# Supplement C. Simulation Details

## C.1 Scalar Contrast Simulation

This simulation isolates a simple design failure mode: a surrogate can be highly accurate for individual prediction while still sending generic uncertainty sampling away from the observations that matter for a prespecified contrast. The estimand is the difference in protocol means between two groups. Each group has $n = 30{,}000$ simulated units, with protocol means 0.351 in group $G_0$ and 0.454 in group $G_1$, giving a protocol contrast of 0.103. The surrogate has high overall accuracy, 0.929, but its error is contrast-aligned: it overpredicts the lower group by 0.060 and is nearly unbiased in the higher group. The resulting surrogate-only contrast is 0.042, with contrast-aligned bias $-0.062$.

The validation policies compare four ways of spending a fixed expected validation budget. Random validation is the design-valid reference. Generic uncertainty validation prioritizes large prediction ambiguity. Target-specific validation uses the recorded-probability variance contribution for the target contrast. The oracle contrast row is an infeasible lower-bound diagnostic that uses full residual information. Budgets range from 100 to 1,600 validations; the main visual contrast uses budget 400 because it is large enough to show the allocation pattern while remaining in the scarce-validation regime.

The reported metrics are Monte Carlo bias, standard deviation, root mean squared error, model-based standard error, CI half-width, empirical coverage, mean validation count, and expected validation spent in the high-influence stratum. The interpretation is design-based: all policies are evaluated under recorded validation probabilities for the same finite simulation population, and efficiency is measured for the prespecified contrast rather than for individual prediction accuracy.

Supplement Figure C.1 displays the simulation mechanism, and Supplement Table C.1 records the compact numerical summary used to read the mechanism.

**Supplement Table C.1. Compact simulation summary.**

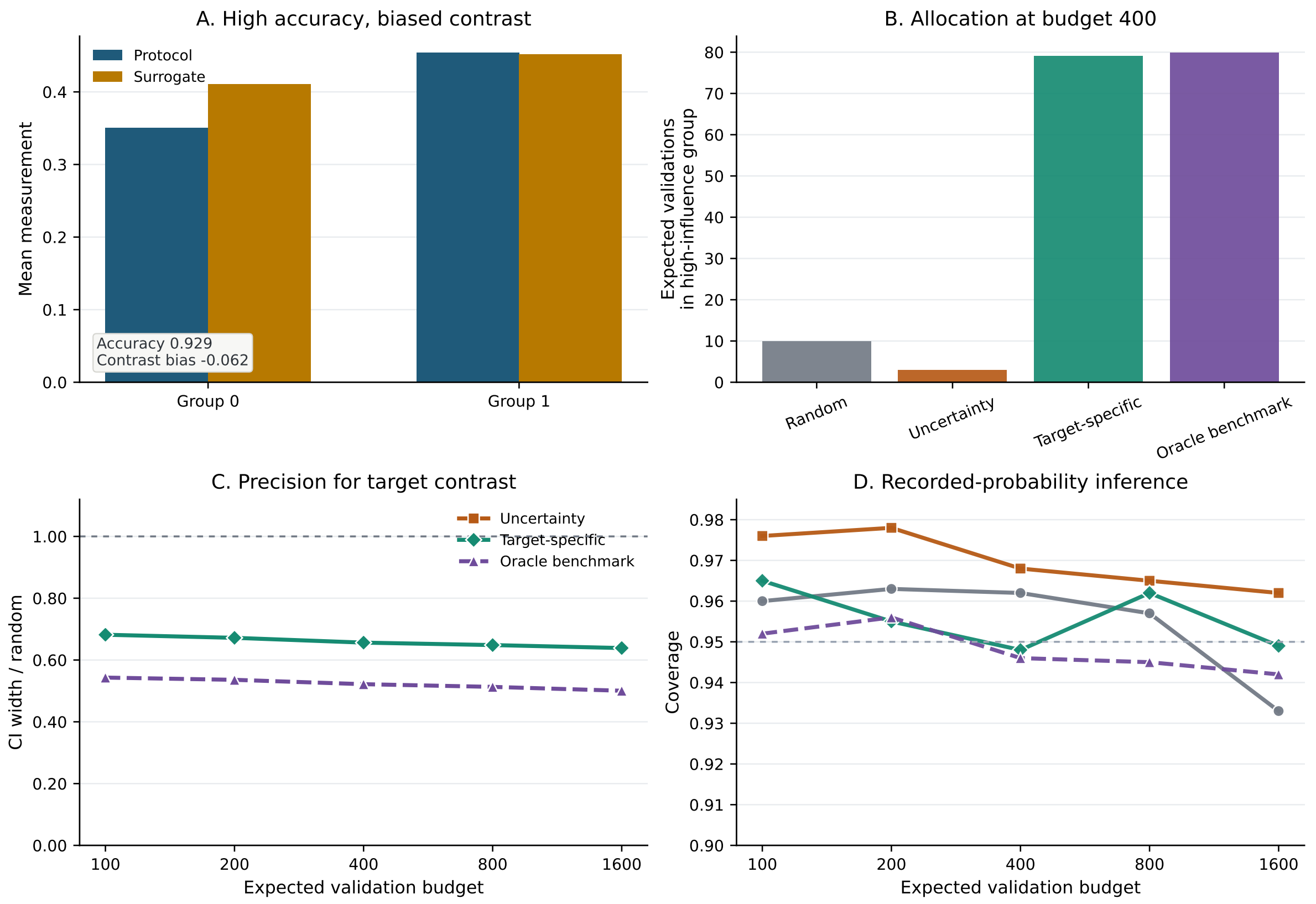


**Supplement Figure C.1.** Simulation evidence that generic prediction uncertainty can be misaligned with a prespecified contrast. The figure visualizes the simulation summarized in Supplement C: high surrogate accuracy does not by itself guarantee small contrast-aligned bias or efficient validation allocation.

| Quantity | Value |
|---|---|
| Surrogate accuracy | 0.929 |
| Protocol contrast | 0.103 |
| Surrogate-only contrast | 0.042 |
| Contrast-aligned bias | -0.062 |
| Budget-400 uncertainty CI ratio | 1.52 |
| Budget-400 target-specific CI ratio | 0.66 |

# Supplement D. A4/LEARN Empirical Details and Robustness

The A4/LEARN empirical supplement is organized around four robustness questions. First, which variables are plausibly first-phase surrogates, and how much archive support do they have? Second, does the APOE4 PET contrast remain stable across nearby Centiloid cutoffs, budgets, and comparator families? Third, do visual-read and cohort diagnostics support the biomedical interpretation without changing the endpoint? Fourth, what feature-support, surrogate-sensitivity, and provenance boundaries should govern a controlled-access retrospective emulation?

## D.1 Feature Timing, Availability, and Leakage Risk

The A4/LEARN surrogate analyses use first-phase archive variables to predict the protocol PET target. Supplement Table D.1 separates these variables from the amyloid PET outcome and records the reviewer-facing timing interpretation. The table is a reproducibility and leakage-defense table for the retrospective validation-design emulation, not a clinical data dictionary.

Terminology in this supplement follows the main text: a surrogate is pre-validation proxy information or a working PET-risk score built from such information. It is not a surrogate endpoint. Amyloid PET remains the protocol measurement that defines the finite-archive target; the surrogate variables are used only for prediction, validation allocation, control-variate construction, and sensitivity diagnostics.

**Compact feature glossary.** PACC denotes the Preclinical Alzheimer Cognitive Composite, a cognitive summary used here as first-phase cognitive support rather than as an endpoint. MMSE is the Mini-Mental State Examination total score. Logical memory summaries measure delayed episodic-memory performance. DSST denotes the Digit Symbol Substitution Test total score, a processing-speed and attention summary. Plasma variables include lower-cost blood biomarker summaries such as GFAP, A$\beta$40/42, NfL, and pTau181; pTau217 is separated because its timing support is mostly but not uniformly pre-PET in this archive. MRI summaries include hippocampal volumes and hippocampal occupancy score, used as structural imaging proxies rather than as PET outcomes. These definitions are intentionally compact because the analysis uses feature families to support validation design, not to make clinical claims about the features themselves.

**Supplement Table D.1. A4/LEARN feature timing and availability.** The PET+APOE set contains 4,460 participants with nonmissing Centiloid PET target and APOE4 group. "Any available" means at least one listed variable in the feature family is nonmissing; "complete available" means all listed variables in the family are nonmissing.

| Feature family | Variables | PET+APOE any | PET+APOE complete | Timing/leakage | Handling |
|---|---|---|---|---|---|
| Demographics | Age, sex, race, education | 4,460 | 4,454 | Baseline/screening subject information; low leakage risk. | Keep in main surrogate. |
| APOE genotype/carrier | APOE genotype / carrier flag | 4,460 | 4,460 | Pre-PET subject information; not outcome leakage, but circularity concern because APOE4 defines the primary contrast. | Keep with conceptual defense and APOE4-excluded sensitivity. |
| Substudy/cohort indicator | Substudy indicator | 4,460 | 4,460 | Design/context variable known before validation; needed for cohort-aware surrogate specification. | Present cohort-unaware and cohort-aware A4-vs-LEARN diagnostics. |
| Cognitive summaries | PACC, MMSE, logical memory, DSST summaries | 1,708 | 1,167 | Archive summaries; some visit-specific variables require timing caution. | Treat as retrospective first-phase archive summaries unless timing is confirmed. |
| Plasma biomarkers | GFAP, A$\beta$40/42, NfL, pTau181 | 1,926 | 1,349 | Lower-cost biomarker surrogates; specimen timing relative to PET needs confirmation. | Use in retrospective design emulation; confirm timing before prospective language. |
| pTau217 biomarkers | first/v6 pTau217 summaries | 1,439 | 1,000 | Biomarker summaries; timing relative to PET needs confirmation. | Report availability; keep outside the main fixed-threshold surrogate unless timing is documented. |
| MRI summaries | bilateral hippocampus, hippocampal occupancy | 1,792 | 1,792 | Lower-cost imaging surrogate; scan timing relative to PET needs confirmation. | Use in retrospective design emulation; confirm timing before prospective language. |
| Amyloid PET protocol measurement | Centiloid, SUVR, CL24 indicator, visual read | 4,460 | 4,460 | Second-phase protocol measurement; high leakage risk if used as surrogate. | Never use in a deployable surrogate score; use only as protocol target or descriptive measurement. |

The main text therefore describes the complete-PET A4 analyses as retrospective design emulations. This framing is deliberately conservative: cross-fitting prevents an individual's own PET value from entering its predicted surrogate risk, but it does not by itself establish that every archive variable would be available before a prospective PET validation decision. The two-wave pilot analysis is the deployable workflow evidence because it learns residual uncertainty from a random PET pilot and uses recorded validation probabilities.

The feature families in Supplement Table D.1 should be read as a layered archive taxonomy.

Demographics, APOE4, and substudy/cohort indicators are baseline or design information with broad support in the PET+APOE analysis set. Cognitive summaries, Roche plasma biomarkers, and MRI summaries are heterogeneous auxiliary archive measurements: useful for retrospective design emulation and model-ladder sensitivity, but not universally observed across all A4/LEARN strata. pTau217 is retained as sensitivity-only because local timing evidence is mostly but not uniformly pre-PET. A prospective scarce-PET validation workflow would restrict the design score to variables documented as available before final PET validation probabilities are assigned, or would learn the score from a random pilot.

## D.2 Cutoff, Budget, and Comparator Robustness

The primary endpoint is Centiloid $\geq 24$. Nearby thresholds 20 and 30 are reported as an operational sensitivity audit, not as alternative primary endpoints. The key pattern is stable: the PET protocol APOE4 contrast remains positive across the three cutoffs, surrogate-only bias is modest near the primary threshold, and integrated cutoff-range validation improves CI width relative to random validation at budget 200.

Supplement Table D.2 gives the three-threshold numerical audit corresponding to the cutoff discussion.

**Supplement Table D.2. APOE4 amyloid PET protocol contrast across Centiloid thresholds.**

| Threshold | PET APOE4 contrast | Surrogate contrast | Surrogate bias | AUC | Random CI half-width | Uncertainty CI ratio | Fixed-24 CI ratio | Process CI ratio |
|---|---|---|---|---|---|---|---|---|
| 20 | 0.326 | 0.327 | +0.002 | 0.882 | 0.097 | 0.899 | 0.865 | 0.861 |
| 24 | 0.334 | 0.347 | +0.012 | 0.908 | 0.090 | 0.905 | 0.868 | 0.859 |
| 30 | 0.318 | 0.349 | +0.031 | 0.948 | 0.081 | 1.008 | 0.949 | 0.918 |

At the primary cutoff $c = 24$, the PET+APOE analysis set has 4,460 participants, the protocol APOE4 contrast is 0.334, the surrogate-only contrast is 0.347, the surrogate bias is +0.012, the threshold-specific AUC is 0.908, and the integrated cutoff-range CI ratio is 0.859 at budget 200.

Oracle ratios are not reported for the cutoff-process simulation because the cutoff-process

output does not contain a matching oracle design; the table leaves those cells missing rather than borrowing an oracle ratio from a different experiment.

Supplement Figure D.1 gives the corresponding cutoff-indexed multiplier-band evidence. It complements the three-panel main Figure 3 by showing the simultaneous-band calculation over the same operational Centiloid range, while keeping the main paper focused on the PET protocol contrast, surrogate bias, and validation-efficiency patterns.

**Budget-curve and baseline placement.** Budget 200 is used in the main displays as a readable mid-range scarce-PET scenario, not as a uniquely selected optimum. The existing budget curves show the same qualitative APOE4 pattern across budgets: generic uncertainty sampling is close to random validation, while target-specific validation is consistently shorter in CI half-width. In the deployable two-wave pilot workflow, the same pattern appears when the final-stage validation probabilities are learned from a random pilot rather than from final-stage PET outcomes.

Supplement Table D.2a records the budget curve, and Supplement Table D.2b records which practical comparators are available in the current evidence package.

**Supplement Table D.2a. APOE4 budget-curve summary.** Ratios are relative to random validation for the fixed-threshold design and to pilot-random validation for the two-wave pilot design. Oracle rows are infeasible lower bounds, not proposed deployable methods.

| Workflow | Budget | Uncertainty CI ratio | Target-specific CI ratio | Oracle lower-bound CI ratio |
|---|---|---|---|---|
| fixed-threshold APOE4 | 50 | 0.979 | 0.917 | 0.812 |
| fixed-threshold APOE4 | 100 | 0.984 | 0.917 | 0.809 |
| fixed-threshold APOE4 | 200 | 0.988 | 0.915 | 0.805 |
| fixed-threshold APOE4 | 400 | 0.987 | 0.911 | 0.793 |
| fixed-threshold APOE4 | 800 | 0.986 | 0.902 | 0.766 |
| two-wave pilot APOE4 | 100 | 0.993 | 0.934 | 0.816 |
| two-wave pilot APOE4 | 200 | 0.991 | 0.919 | 0.808 |
| two-wave pilot APOE4 | 400 | 0.994 | 0.918 | 0.801 |
| two-wave pilot APOE4 | 800 | 0.989 | 0.910 | 0.778 |

The available empirical baselines in the current analysis package are random validation, generic uncertainty validation, one-dimensional balanced validation at budget 200 for the primary APOE4, age-slope, and A4-vs-LEARN diagnostic targets, target-specific validation, two-wave pilot-learned target-specific validation, and oracle lower-bound validation. The broader balanced follow-up adds age-quartile and cohort-balanced budget-200 rows to the existing APOE4-balanced row. Crossed balanced designs have feasibility count checks only; cutoff-aware balanced validation is still not available in the current cutoff-process output.

**Supplement Table D.2b. Comparator availability and budget-200 interpretation boundary.** CI and RMSE ratios are relative to random validation under the same independent Bernoulli expected-budget-200 convention. Crossed feasibility rows are count checks and are not performance evidence.

| Comparator family | Current support | Budget-200 ratios where available | Supported interpretation | Present scope |
|---|---|---|---|---|
| Random validation | Reported in fixed-threshold, cutoff, and pilot rows | Reference ratio 1.000 | Design-valid reference under recorded probabilities | Not an efficiency optimum |
| Generic uncertainty validation | Reported in the main diagnostics and budget table | APOE4 CI ratio 0.980; age-slope CI ratio 1.000; A4-vs-LEARN CI ratio 0.980 | Common practitioner heuristic for ambiguous predictions | Heuristic comparator rather than closest theoretical benchmark |
| APOE4-balanced validation | Available for the primary APOE4 budget-200 display | CI ratio 0.923; RMSE ratio 0.938; coverage 0.956 | Simple balancing is nearly enough for the scalar APOE4 contrast; target-specific CI ratio is 0.914 | Does not cover age-band, cohort, crossed, or cutoff-aware balancing |
| Age-quartile balanced validation | Available for the age-slope budget-200 display | CI ratio 1.001; RMSE ratio 1.012; coverage 0.955 | Simple age-quartile balancing does not reproduce the target-specific age-slope gain; target-specific CI ratio is 0.806 | Budget-200 one-dimensional check only, not a full age-balanced grid across budgets |
| Cohort-balanced validation | Available for the A4-vs-LEARN budget-200 diagnostic display | CI ratio 0.940; RMSE ratio 1.001; coverage 0.928 | Cohort balancing helps CI width but remains diagnostic; target-specific CI and RMSE ratios are 0.916 and 0.946 | Not a biological or causal cohort claim |
| Target-specific and pilot-learned target-specific validation | Reported in main displays and Supplement Table D.2a | Target-specific ratios vary by target and budget; pilot rows use pilot-random as reference | Residual-uncertainty targeting can be evaluated under the same recorded-probability convention, including a deployable pilot workflow | Crossed, cutoff-balanced, and active/PPI/OPAL-style rules are not evaluated here |
| Oracle lower-bound validation | Reported as infeasible lower-bound diagnostics | APOE4 budget-200 CI ratio 0.804; age-slope CI ratio 0.699; A4-vs-LEARN CI ratio 0.733 | Shows the best-case residual-information benchmark | Not deployable and not a recommended design |
| Crossed stratified validation | Feasibility count checks only for APOE4 x age quartile, APOE4 x substudy, and substudy x age quartile | No performance ratios; minimum cell counts 329, 123, and 99, respectively | Candidate crossed grids are count-feasible under a budget-200 equal-cell thought experiment | Performance comparison is not evaluated here |
| Cutoff-aware balanced comparator | Not available in the current cutoff-process output | No performance ratios | Cutoff-process evidence supports the recorded-probability target-specific and process-band calculations | Purpose-built cutoff-balanced performance is not evaluated here |

The information sets differ across rows. Full-PET retrospective emulations use the complete archive only to hide outcomes, compare validation policies, and score what each policy would have achieved for the finite-archive protocol target. Oracle lower-bound rows use full-protocol residual information and are evaluation-only; they are not deployable validation designs. The two-wave pilot rows are the deployable counterpart: the first wave is random PET validation used to learn residual protocol uncertainty, the final-stage probabilities are

assigned from first-phase and pilot/history information, and each final probability is recorded before the corresponding final PET outcome contributes to the estimator.

## D.3 PET Visual-Read and Cohort Diagnostics

The visual PET read is used as a biomedical agreement diagnostic for the operational Centiloid endpoint. It is not treated as a replacement truth variable. In 4,492 paired rows, visual PET positivity and Centiloid $\geq 24$ agree in 92.2% of participants, with Cohen's kappa 0.821.

Supplement Table D.3 reports the corresponding visual-read agreement audit.

**Supplement Table D.3. Visual PET agreement with** Centiloid $\geq 24$**.**

| Visual variable | Paired n | CL24+ n | Visual+ n | Agree | Kappa | Sens | Spec | PPV | NPV | Note |
|---|---|---|---|---|---|---|---|---|---|---|
| processed visual positive | 4,492 | 1,528 | 1,329 | 0.922 | 0.821 | 0.821 | 0.975 | 0.944 | 0.913 | Usable paired diagnostic |

Discordant rows should be interpreted as expected read-threshold differences rather than data errors. This diagnostic supports the clinical plausibility of the Centiloid target while preserving the continuous Centiloid-derived endpoint as the protocol measurement used in the main analysis.

**Stratified visual-read discordance.** The overall visual-read agreement is similar across major APOE4, substudy, and age strata. The stratified audit is descriptive: it checks whether the CL24 endpoint has obvious biomedical discordance patterns, but it does not replace the Centiloid endpoint or define a new visual-read target analysis.

Supplement Table D.3a gives the stratified discordance audit.

**Supplement Table D.3a. Stratified visual-read discordance with** Centiloid $\geq 24$**.** The discordance direction column reports visual-positive/CL24-negative rows versus visual-negative/CL24-positive rows.

| Stratifier | Level | Paired n | Agreement | Discordance | Direction counts |
|---|---|---|---|---|---|
| APOE4 | APOE4+ | 1,563 | 0.915 | 0.085 | 15 vs 118 |
| APOE4 | APOE4- | 2,897 | 0.926 | 0.074 | 60 vs 155 |
| Substudy | A4 | 1,169 | 0.948 | 0.052 | 61 vs 0 |
| Substudy | LEARN | 538 | 0.926 | 0.074 | 0 vs 40 |
| Substudy | Screening/SF | 2,785 | 0.911 | 0.089 | 14 vs 234 |
| Age band | 65-69 | 2,099 | 0.912 | 0.088 | 50 vs 134 |
| Age band | 70-74 | 1,446 | 0.928 | 0.072 | 16 vs 88 |
| Age band | 75+ | 945 | 0.935 | 0.065 | 9 vs 52 |

The A4 stratum is a useful reminder that visual read and CL24 encode different operational thresholds: all paired A4 visual reads are positive in this processed audit, while CL24 positivity is 94.8%. This is why the visual read is used as an agreement audit rather than as the primary validation endpoint.

**Within-cohort APOE4 descriptive audit.** The primary APOE4 contrast remains the finite-archive PET protocol contrast used in the main validation-design analysis. Within-cohort APOE4 contrasts are reported only as a descriptive biomedical robustness audit because the substudy strata have different enrichment and positivity structures.

Supplement Table D.3b reports these within-cohort APOE4 descriptive rows.

**Supplement Table D.3b. Within-cohort APOE4 PET positivity audit.**

| Substudy | n | $n_+$ | $n_-$ | PET+ rate $G_+$ | PET+ rate $G_-$ | Diff. | Interpretation |
|---|---|---|---|---|---|---|---|
| A4 | 1,169 | 689 | 480 | 0.983 | 0.898 | 0.085 | Interpretable descriptive sensitivity, but positivity is near-saturated. |
| LEARN | 536 | 123 | 413 | 0.098 | 0.068 | 0.030 | Limited by near-zero PET positivity. |
| Screening/SF | 2,755 | 751 | 2,004 | 0.246 | 0.096 | 0.151 | Most interpretable within-cohort descriptive contrast. |

$G_+$ and $G_-$ denote APOE4 carrier and noncarrier groups; Diff. is carrier minus noncarrier PET positivity.

These within-cohort rows support the biomedical plausibility of an APOE4 PET difference without turning the analysis into a new cohort-specific endpoint study. A4 enrichment compresses the within-A4 contrast; LEARN has low PET positivity; Screening/SF is more interpretable descriptively but is not the randomized A4 cohort.

## D.4 Feature Support, Surrogate Sensitivity, and Provenance Boundary

The missingness audit stratifies rows by PET status, APOE4 status, substudy, age band, and feature family. It is a stratum-level diagnostic, not a replacement for the feature-family availability counts in Supplement Table D.1. The percentages in Supplement Table D.4 are medians across 55 strata, and the total count is the sum of stratum sizes used in the audit rather than a unique participant denominator. A median of zero therefore means that at least half of the audited strata have no support for that feature family; it does not mean that the feature family is universally absent.

**Supplement Table D.4. Median feature-family support across stratified A4/LEARN audit rows.**

| Feature family | Strata | Total stratum n | Median all-complete pct | Median any-observed pct | Median mean nonmissing pct |
|---|---|---|---|---|---|
| APOE | 55 | 6,945 | 1.000 | 1.000 | 1.000 |
| Cognition | 55 | 6,945 | 0.000 | 0.000 | 0.000 |
| Demographics | 55 | 6,945 | 1.000 | 1.000 | 1.000 |
| MRI | 55 | 6,945 | 0.255 | 0.255 | 0.255 |
| PET protocol | 55 | 6,945 | 1.000 | 1.000 | 1.000 |
| Plasma | 55 | 6,945 | 0.000 | 0.524 | 0.292 |

Demographics, APOE, and the PET protocol fields have broad support in the relevant analysis strata. Plasma and MRI are useful sensitivity families but have visibly stratified support, so the manuscript treats them as retrospective archive measurements unless timing and availability are confirmed for a prospective validation workflow. The cognition row should be read with the same stratum-level caution: Supplement Table D.1 shows that cognitive summaries are available for 1,708 PET+APOE participants with at least one listed cognitive variable and for 1,167 participants with all listed cognitive variables. The zero medians in Supplement Table D.4 arise because cognitive summaries are concentrated in a subset of the stratified archive cells, especially after PET status, APOE4 status, substudy, and age-band stratification.

**Initial model-ladder and calibration diagnostics.** The initial model ladder reports AUC, Brier score, calibration, contrast-aligned residual bias, and validation-efficiency diagnostics. This table is a Sprint 0 credibility audit rather than a final model-development exercise.

Supplement Table D.5 gives the transparent and sensitivity surrogate-model diagnostics.

**Supplement Table D.5. Transparent and sensitivity surrogate-model diagnostics.**

| Model | Role | AUC | Brier | Calib. | ECE | APOE4 resid. | Target CI | Oracle CI |
|---|---|---|---|---|---|---|---|---|
| Clinical + APOE | main | 0.719 | 0.211 | 0.996 | 0.134 | +0.018 | 0.927 | 0.846 |
| Clinical + APOE + plasma | sensitivity | 0.759 | 0.196 | 0.979 | 0.125 | +0.018 | 0.923 | 0.822 |
| Clinical + APOE + MRI | sensitivity | 0.749 | 0.200 | 1.014 | 0.155 | +0.018 | 0.925 | 0.828 |
| Clinical + APOE + plasma + MRI | sensitivity | 0.780 | 0.186 | 1.004 | 0.124 | +0.018 | 0.917 | 0.805 |
| Cohort-aware full | sensitivity | 0.906 | 0.095 | 0.992 | 0.067 | +0.012 | 0.880 | 0.632 |
| APOE-excluded full | sensitivity | 0.724 | 0.206 | 1.005 | 0.144 | -0.265 | 0.952 | 0.875 |

Calib. is calibration slope; ECE is decile expected calibration error; CI columns are half-width ratios relative to random validation under the same budget convention.

AUC alone is not used as evidence of measurement validity. The calibration and contrast-residual columns are included because the paper's estimand is a scientific contrast, not individual PET prediction.

The model ladder also records the model-agnostic role of the framework. A transparent calibrated surrogate is used for the primary analysis because it is easier to audit for timing, leakage, and biomedical interpretation. Richer calibrated learners or ensemble scores can be treated as sensitivity analyses when their timing and missingness are documented, but they are working first-phase scores rather than standalone clinical PET prediction tools. Across surrogate specifications, design validity is supplied by the recorded validation probabilities; surrogate quality affects efficiency and is diagnosed through calibration, Brier score, contrast-aligned residual bias, and validation-design metrics rather than AUC alone.

Supplement Table D.5a gives the compact model-agnostic surrogate sensitivity.

**Supplement Table D.5a. Compact model-agnostic surrogate sensitivity.** All rows use cross-fitted PET-risk predictions from the processed analysis-ready archive.

| Model | Role | AUC | Brier | APOE4 resid. | Uncert. CI | Target CI | Timing note |
|---|---|---|---|---|---|---|---|
| Logistic clinical+APOE | transparent baseline | 0.719 | 0.211 | +0.019 | 1.006 | 0.928 | No pTau217; transparent baseline. |
| Penalized clinical+APOE | calibrated sensitivity | 0.721 | 0.192 | -0.015 | 0.969 | 0.921 | No pTau217; held-out calibration. |
| Penalized full safe | safe-feature sensitivity | 0.778 | 0.169 | -0.015 | 0.966 | 0.921 | Plasma and MRI archive features; pTau217 excluded. |
| Penalized cohort-aware | cohort-aware sensitivity | 0.906 | 0.088 | -0.003 | 0.920 | 0.887 | Cohort indicator included as measurement context. |
| Random forest full safe | flexible sensitivity | 0.901 | 0.091 | -0.021 | 0.932 | 0.896 | Flexible calibrated learner; pTau217 excluded. |
| Penalized full + pTau217 | pTau217 timing sensitivity | 0.851 | 0.130 | +0.015 | 1.098 | 1.064 | Sensitivity only; timing mostly but not uniformly pre-PET. |

Uncert. and Target CI are APOE4 CI half-width ratios for generic uncertainty and target-specific validation, respectively.

The no-pTau217 rows show that changing from a transparent score to calibrated penalized or flexible scores does not remove the APOE4 target-specific validation advantage. The pTau217 row is a cautionary sensitivity: stronger prediction summaries do not automatically improve validation design for the APOE4 contrast, and timing uncertainty keeps pTau217 outside the main feature set.

**APOE4-excluded surrogate sensitivity.** Because APOE4 defines the primary contrast and can also enter the PET risk surrogate, the APOE4-excluded sensitivity keeps APOE4 as the prespecified target-defining group but removes APOE4 from the surrogate prediction model. This prevents the surrogate score from directly encoding the APOE4 contrast.

Supplement Table D.6 reports the APOE4-included and APOE4-excluded comparison.

**Supplement Table D.6. APOE4-included versus APOE4-excluded surrogate sensitivity.**

| Feature specification | Surrogate AUC | PET protocol contrast | Surrogate-only contrast | Surrogate bias | Budget-200 target-specific CI ratio | Budget-200 target-specific RMSE ratio |
|---|---|---|---|---|---|---|
| Surrogate includes APOE4 | 0.780 | 0.334 | 0.352 | +0.018 | 0.921 | 0.919 |
| Surrogate excludes APOE4 | 0.724 | 0.334 | 0.069 | -0.265 | 0.952 | 0.961 |

Removing APOE4 from the surrogate materially compresses the surrogate-only APOE4

contrast, which strengthens the measurement-warning message. Target-specific validation remains valid because APOE4 is used as a prespecified influence weight for the estimand and final inference uses recorded validation probabilities.

**Provenance and prospective-use cautions.** The A4/LEARN analyses above are reported as retrospective design emulations using the local controlled-access archive. The user-confirmed source identity is the A4 and LEARN Study Data Portal, A4StudyData.org, release A4_LEARN_Data_Release_2024-07; controlled-access A4/LEARN access was obtained on 2026-05-09 at 09:03. Controlled-access individual-level files are not redistributed. Data-access, workflow, and ACC-style checklist instructions are summarized in Supplement E; full checksum and source-archive provenance records are retained locally and are not redistributed with the public arXiv package.

Plasma specimen timing and MRI scan timing relative to amyloid PET remain conservative interpretation items. Local date-field evidence supports Roche plasma and volumetric MRI as pre-PET among overlapping participants, but their availability is still study-structured rather than universal. These limitations do not affect the internal validity of the retrospective design emulations, but they limit claims that every surrogate feature would be available before PET in a prospective validation workflow.

# Supplement E. Reproducibility and Package Notes

The reproducibility materials separate three audiences. Authorized reviewers with A4/LEARN access can rebuild the controlled-access empirical outputs. Public readers without A4/LEARN access can run the shareable simulation smoke test. The public source package documents which derived tables and figures are shareable, while the raw A4/LEARN files remain nonredistributable.

Supplement Table E.1 collects the review-facing pointers for these audiences.

The controlled-access A4/LEARN workflow is run from the package root with:

```
bash scripts/run_a4_main_outputs.sh
```

The public mechanics smoke test is:

```
python3 scripts/30_method_simulations.py
```

This smoke test exercises the recorded-probability scalar-contrast simulation workflow and reproduces the shareable simulation evidence in Supplement C. It does not reproduce the A4/LEARN PET estimates, endpoint audits, or controlled-access figures.

**Supplement Table E.1. Compact review-facing pointers.**

| Purpose | Pointer | Role |
|---|---|---|
| Start guide | `docs/aer_start_here_blinded_20260609.md` | Reviewer-facing entry point for controlled-access and public workflows. |
| Blinded manifest | `docs/blinded_reproducibility_manifest_20260608.md` | Lists package files, derived outputs, and controlled-access boundaries. |
| ACC checklist | `docs/jasa_acc_reproducibility_checklist_blinded_20260609.md` | Journal-style reproducibility checklist. |
| Controlled-access workflow | `scripts/run_a4_main_outputs.sh` | Runs the A4/LEARN empirical pipeline after authorized local data staging. |
| Public smoke test | `scripts/30_method_simulations.py` | Rebuild the shareable scalar recorded-probability simulation. |
| Simulation aggregates | `results/method_simulation_design_summary.csv`; `results/method_simulation_bias.csv` | Machine-readable summaries behind Supplement C. |

Simulation and validation-design scripts set their random seeds internally when Monte Carlo sampling or design emulation is used. The A4/LEARN aggregate CSVs underlying Supplement D are named in the blinded manifest rather than repeated throughout this PDF. Validation-design inference uses recorded validation probabilities produced by the design scripts; probability recording is part of the estimand and variance calculation rather than a display-only diagnostic.

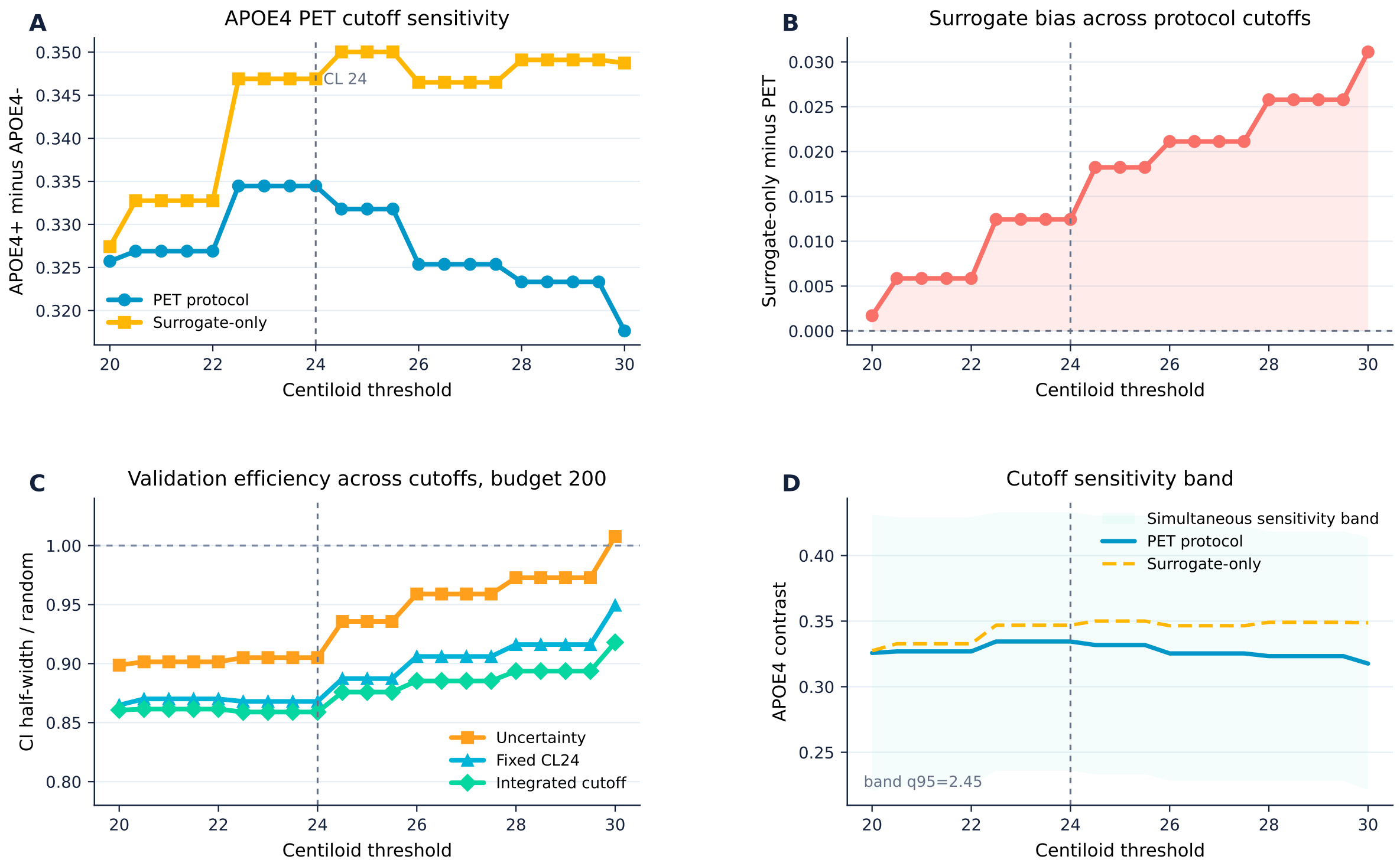


**Supplement Figure D.1.** Supplement-facing cutoff-indexed multiplier-band evidence over the operational Centiloid range. The display supports the recorded-probability process discussion in Supplement B.5 and the cutoff sensitivity summary in Supplement D.2.



\end{document}